\documentclass[12pt]{iopart}
\usepackage{iopams}
\usepackage{setstack}
\usepackage{graphicx}
\usepackage{epsfig}

\usepackage{epstopdf}
\usepackage{times}
\usepackage{txfonts}
\usepackage{xcolor}

\usepackage{bm}        
\usepackage{amssymb}   
\usepackage{verbatim}  
\usepackage{epstopdf}

\newcommand{\beq}{\begin{equation}}
\newcommand{\eeq}{\end{equation}}

\newcommand{\commentout}[1]{{}}
\newcommand{\idx}{{l}}

\newcommand{\<}{\langle}
\renewcommand{\>}{\rangle}
\renewcommand{\(}{\left(}
\renewcommand{\)}{\right)}
\renewcommand{\[}{\left[}
\renewcommand{\]}{\right]}
\newcommand{\eq}[1]{Eq.~(\ref{#1})}

\newcommand{\half}{\hbox{$1\over2$}}

\begin{document}

\title[A two-species Bose-Einstein condensate in an optical lattice]{Fragmentation, domain formation and atom number fluctuations of a two-species Bose-Einstein condensate in an optical lattice}
\author{Uttam Shrestha}
\address{Department of Physics and Astronomy, University of California, Irvine, California 92697-4575, USA}
\address{LENS, Universit\`a di Firenze, via Nello Carrara 1, 50019 Sesto Fiorentino, Firenze, Italy}
\author{Janne Ruostekoski}
\address{School of Mathematics, University of Southampton,
Southampton, SO17 1BJ, UK}
\date{\today}

\begin{abstract}
We theoretically study the loading of a two-species Bose-Einstein condensate to an optical lattice in a tightly-confined one-dimensional trap. Due to quantum fluctuations the relative inter and intra species phase coherence between the atoms and the on-site atom number fluctuations are reduced in the miscible regime. For the immiscible case the fluctuations are enhanced and the atoms form metastable interleaved spatially separated domains where the domain length and its fluctuations are affected by quantum fluctuations.
\end{abstract}

\maketitle
\section{Introduction}

Two-species atomic Bose-Einstein condensates (BECs) exhibit notably richer dynamical phenomena than single-species BECs. The inter-species interactions between the two components affect the nonlinear dynamics of the individual BECs and the two-species mixture may be in a miscible or immiscible phase \cite{Hall,Papp,Esry,Timmermans,Law,Sinatra99,Kasamatsu,Ruostekoski_Dutton}, exhibiting, e.g., spin \cite{LEW02} and shock waves \cite{Dutton_Science_2001}, vector solitons \cite{Dutton_Science_2001,RUO01, Busch_PRL_2001,SAV03,Anderson_PRL_2001,Becker_Nature_2008,Shrestha,Hamner,Yan} and  other topological defects and textures \cite{Kasamatsudefect}. In optical lattices
a bosonic two-component mixture has attracted an increasing experimental interest, e.g., in controlled collisions and multi-particle entanglement \cite{MAN03}, in mixing of $^{87}$Rb and $^{41}$K in 3D lattices \cite{CAT07}, in super-exchange interactions \cite{superex}, in spin-gradient thermometry \cite{spinther}, and in sub-shot-noise quantum interferometry \cite{Gross_Nature_2010}.

In this paper we study both numerically and analytically the effects of quantum and thermal fluctuations on a two-species BEC when the condensates are confined in an optical lattice in a highly-elongated 1D trap and the lattice potential is slowly turned up. In single-species bosonic atomic gases the interplay between enhanced quantum fluctuations in an optical lattice and the repulsive inter-atomic interactions has experimentally been shown to result in strongly reduced atom number fluctuations and the loss of phase coherence between the atoms in different lattice sites \cite{ORZ01,Gerbier_PRL_2006,Li_PRL_2007,Stebby_PRL_2007,Esteve_Nature_2008,Gemelke_Nature_2009,Bakr_Science_2010,Sherson_Nature_2010,Gross_PRA_2011}. The reduction in atom number fluctuations has been exploited in preparation of spin-squeezed states \cite{Esteve_Nature_2008} that are suitable for quantum-enhanced interferometry. The phase separation dynamics of a two-species harmonically-trapped BEC was experimentally observed in two immiscible hyperfine components of $^{87}$Rb \cite{Hall}, due to the long lifetimes of the two-fluid system that results from a fortuitous cancellation of the scattering lengths \cite{VOG00}. Controllable spatial separation dynamics has recently been observed in a mixture of $^{85}$Rb and $^{87}$Rb atomic BECs by tuning the inter-species interactions with a magnetic Feshbach resonance \cite{Papp} and by introducing a linear electromagnetic coupling between the two internal states \cite{obert11} that creates an effective dressed state description for the two components \cite{Blakie99,Jenkins}.

In our simulations the atoms are initially confined in a shallow lattice and we continuously turn up the lattice potential. The enhanced effective interactions result from the reduced hopping amplitude of the atoms along the lattice and quantum fluctuations become more dominant in a deep lattice. We calculate the on-site atom number fluctuations in individual lattice sites and the inter and intra species relative phase coherence between the atoms in different sites. The numerical results are compared with the analytically calculated values that we derive in the appendix using the Bogoliubov theory. Even in the miscible regime of the two-species BEC system quantum fluctuations eventually destroy the long-range coherence of the atoms along the lattice, fragmenting the condensates. The inter and intra-species relative phase coherence between the atoms even in the adjacent sites is notably reduced in deep lattices, but the inter-species coherence remains higher close to the onset of the phase separation instability. We find that the repulsive inter-species interactions increase the inter-species relative phase coherence, but have only a weak effect on the intra-species coherence and the on-site atom number fluctuations. The coherence typically stabilizes to a non-vanishing finite value after the lattice ramping and we evaluate its spatial dependence along the lattice, demonstrating a clearly reduced spatial coherence length of the system. In the dynamically unstable regime we find considerably enhanced atom number fluctuations, stronger loss of phase coherence, and the spontaneous formation of metastable configurations of interleaved domains of the two spatially-separated components. We calculate quantum mechanical expectation values and uncertainties of the domain length and find that they depend on the strength of quantum fluctuations, deviating from the classical mean-field values.

The experimentally observed phase separation dynamics of Ref.~\cite{Papp} in the uniform space was theoretically studied using the classical mean-field theory in Ref.~\cite{Ronen}. A multiorbital wavefunction analysis was employed in Ref.~\cite{Alon} to demonstrate that stronger inter-species interactions lead to a shorter domain length. There has been an increasing interest in the experimental studies of spontaneous symmetry breaking and pattern formation also in other ultra-cold atomic systems \cite{spinorpattern1,spinorpattern2} and the two-species condensate with a coupling between the two internal states \cite{obert11} has been proposed as a system to study the Kibble-Zurek defect formation mechanism in phase transitions \cite{LEE09,Zurek}.

In order to perform efficiently the numerical simulations in a lattice, we develop an approximate method to describe the non-equilibrium dynamics of a two-species condensate mixture that is based on the truncated Wigner approximation (TWA) \cite{Drummond_EPhysLett_1993,Steel_PRA_1998,Sinatra_JPhysB_2002,Isella_PRA_2006,Blakie_AdvInPhys_2008,POL10,Martin_PRL_2010} of the stochastic phase space dynamics. The two-species BEC equilibrium state is solved within the classical Bogoliubov approximation where the excitations are evaluated in the tight-binding approximation of the uniform two-species lattice Hamiltonian. The amplitudes of the Bogoliubov phonon modes are then stochastically sampled according to a probability distribution given by the Wigner distributions of the ideal harmonic oscillators, as in the single-component TWA approaches \cite{Isella_PRA_2006}. Each stochastic realization of the initial state is propagated in time according to the classical mean-field dynamics, so that individual stochastic trajectories represent potential outcomes of single experimental runs and quantum mechanical expectation values and fluctuations are calculated from the ensemble averages of the stochastic dynamics. One of the advantages of the approximate two-species model is its simplicity and the possibility to calculate analytic solutions to the initial state mode functions. The analytic approach to linearized excitations can also be used to calculate approximate ground state properties of the two-species system. We use this in Appendix to evaluate the intra-species relative phase coherence and the on-site atom number fluctuations.

\section{Theoretical Model}

\subsection{Classical mean-field equation}

We assume that a two-species BEC is in a tightly-confined highly-elongated 1D trap, so that any density fluctuations along the radial direction perpendicular to the trap axis can approximately be ignored. Along the axial direction the atoms experience an optical lattice potential that is deep enough so that the atoms can be described in the tight-binding approximation in which case one trap mode per lattice site is included in the dynamics. The classical mean-field model then follows our previous descriptions \cite{Ruostekoski_Dutton,Shrestha} and the equation that governs the dynamics of two component BEC is the two-component discrete nonlinear Schr\"odinger equation (TCDNLSE),
\begin{equation}
i{d\over d t}\psi_{n}^{(j)}=-J_j(\psi_{n+1}^{(j)}+\psi_{n-1}^{(j)})+\sum_{k=1}^{2}\chi_{jk} |\psi_{n}^{(k)}|^2\psi_{n}^{(j)}\,,
\label{E1}
\end{equation}
where $J_{j}$ ($J_{j}>0$) and $\psi_{n}^{(j)}$ denote the nearest-neighbour hopping amplitudes and the wavefunction amplitude at the lattice site $n$ of the atoms of species $j$ ($j=1,2$), respectively.
The nonlinearities are given by the interaction coefficients $\chi_{jk}$ $(j,k=1,2)$ which are proportional to the onsite atom-atom interaction strengths and to the overlap of the lowest vibrational state wave functions $\phi_{n}^{(j)}$ (the Wannier functions) of the two species in a given lattice site, i.e.,
\beq
\chi_{jj}\simeq
\frac{4\pi\hbar N_j a_{jj}}{m_j} \int d^3\vec r\, |\phi_{n}^{(j)}(\vec r)|^4 \, ,
\label{E2}
\eeq
and
\beq
\chi_{jk}\simeq \frac{2\pi\hbar\sqrt{N_iN_j} a_{jk}}{\mu} \int d^3\vec r\,|\phi_{n}^{(j)}(\vec r)|^2 |\phi_{n}^{(k)}(\vec r)|^2
\label{E3}
\eeq
for $j\neq k$. The inter and intra species scattering lengths are denoted by $a_{jk}$ and $a_{jj}$, respectively, and $N_j$ is the atom number of the species $j$. Here the reduced atomic mass $\mu$ is given in terms of the atomic  mass of the $j^{\rm th}$ component $m_j$ as
\beq
\mu=\frac{m_1m_2}{(m_1+m_2)}\, .
\label{E4}
\eeq
Since the number of atoms in each species is a conserved quantity, in the following we use the normalization
\beq
\sum_{n=1}^{L} |\psi_{n}^{(j)}|^2=1\,,
\label{E5}
\eeq
where $L$ denotes the number of lattice sites.

\subsection{Parameter regimes of the simulations}\label{parameters}

We study the quantum dynamics of a two-species BEC in an optical lattice. The initial state of the TWA simulations is generated by calculating the classical Bogoliubov modes whose amplitudes are sampled stochastically. The time-evolution for each stochastic realization then follows TCDNLSE of \eq{E1}. The simulations involve a large parameter space. There are three nonlinearities $\chi_{11}$, $\chi_{22}$, and $\chi_{12}$ in \eq{E1} and the hopping amplitudes $J_k$ for the two species may differ. In addition, the two components can be moving with the different carrier momenta $p_k$. The stochastic initial state fixes the atom numbers $N_1$ and $N_2$ (for given $\chi_{11}$, $\chi_{22}$) and we can also vary the number of lattice sites $L$. The lattice potentials of the two components may be shifted with respect to each other and the two species could also experience different radial confinements.
In the finite temperature examples we also vary the initial temperature $T$. In the following we will demonstrate how a simple analytic description for the initial state of the TWA simulations in terms of the Bogoliubov modes may be obtained whenever the two atom currents are equal, i.e., for $J_1\sin (p_1)=J_2 \sin (p_2)$. In order to demonstrate some basic effects of the two-species quantum dynamics in a lattice, we concentrate on a simple set of parameter values for which the mode functions have especially compact analytic expressions. In all the numerical simulations we consider condensates with zero centre-of-mass momenta $p_1=p_2=0$ and
\beq
J_1=J_2,\quad \chi_{11}=\chi_{22},\quad N_1=N_2\,.
\eeq
We show that the particular choice for the set of parameters is by no means necessary, however, and that the general formalism with the analytic initial state derivation is more general. For the selected parameter set we may investigate the main physical phenomena of the two-species lattice dynamics: the effect of the phase separation dynamics can be controlled by the ratio $\chi_{12}/\chi_{11}$, quantum fluctuations and nonlinearity by $\chi_{11}/N_{1}^2$ and $\chi_{11}/J_{1}$.

\subsection{Classical Bogoliubov theory and stability analysis}

We can find steady-state solutions for the TCDNLSE~(\ref{E1}) that represent propagating plane waves \cite{Ruostekoski_Dutton,Shrestha}
\beq
\psi_n ^{(j)}=\,\frac{1}{\sqrt{L}}e^{i(p_j n- \omega_j t)}\, ,
\label{E6}
\eeq
where $\omega_j$ is given by
\beq
\omega_j=-2 J_j\cos p_j+ \sum_{k=1}^{2}\Delta_{jk}, \quad {{\Delta}} _{jk}\equiv{\chi_{jk}\over L}\, .
\label{E7}
\eeq
The carrier wave momenta, $p_j$ are quantized according to
$p_j =P_j \frac{2 \pi}{L} $
where  $P_j$ is an integer that takes value in the interval $[-\frac{L}{2},\frac{L}{2})$.

The linear stability analysis of the steady-state solution was performed in Ref.~\cite{Ruostekoski_Dutton}
using the classical Bogoliubov expansion. In the Bogoliubov approach the wavefunctions for each component in Eq.~(\ref{E1}) are written as
\begin{equation}
 \psi_n^{(j)}= \frac{1}{\sqrt{L}}(1+u_q^{(j)}e^{iq n}-[v_q^{(j)}]^*e^{-iq n}) e^{i(p_j n-\omega_jt)}\,.
\label{E9}
\end{equation}
In the limit of weak perturbations, the system of equations for $u_q^{(j)}$ and $v_q^{(j)}$ may be
expressed as an eigenvalue problem
\begin{equation} \label{E10}
i{d\over dt} \xi_q= \sigma {\mathfrak M}_q \xi_q,\quad  \xi_q=\left(
\begin{array}{c}
u_q^{(1)} \\
v_q^{(1)}\\
u_q^{(2)}\\
v_q^{(2)}
\end{array}
\right),\quad \sigma=\left(
\begin{array}{cc}
\sigma_z & 0 \\
0 & \sigma_z
\end{array}
\right) \,,
\end{equation}
where $\sigma_z$ denotes the $2\times2$ Pauli spin matrix. The
elements of the $4\times4$ matrix ${\mathfrak M}_q$ are obtained from the Bogoliubov linearization
procedure \cite{Ruostekoski_Dutton}. The quasimomenta $q=\frac{2\pi Q}{L}$ may be defined such that $Q$ takes integer values except zero in the range  $[-\frac{L}{2},\frac{L}{2})$.
The eigenvalues of ${\mathfrak M}_q$ correspond to the normal mode (excitation) frequencies $\Omega_q$ of the system that have simple analytic expressions when two BECs have the same atomic currents \cite{Ruostekoski_Dutton}($
 J_1\sin (p_1)=J_2 \sin (p_2)$).
In that case we obtain~\cite{Ruostekoski_Dutton,Shrestha}
\begin{equation}
\Omega_q=t_1\pm\sqrt{\half(t_2\pm t_3)}\, ,
\label{E13}
\end{equation}
where
\beq
t_1 = 2J_1 \sin(p_1)\sin(q)\, ,
\label{E14}
\eeq
represents a Doppler shift term of the excitation frequencies due to the superfluid flow,
\beq
 t_2 = \nu_{1,q}^2+\nu_{2,q}^2\, ,
\label{E15}
\eeq
\beq
t_3 = \sqrt{(\nu_{1,q}^2-\nu_{2,q}^2)^2 +16\epsilon_{1,q}\epsilon_{2,q}\Delta_{12}^2\cos p_1\cos p_2 }\, ,
\label{E16}
\eeq
are defined in terms of the single-condensate normal mode frequencies $\nu_{j,q}$
\beq
\nu_{j,q}^2=\epsilon_{j,q}\cos(p_j) [\epsilon_{j,q}\cos(p_j)+{2\Delta_{jj}}]\,,
\label{E17}
\eeq
and
\beq
\epsilon_{j,q}=4J_j \sin^2(q/2)\, ,
\label{E18}
\eeq
is the spectrum of an ideal, non-moving BEC. The flow is stable if the frequencies in Eq.~(\ref{E13}) are real for all $q\ne0$, otherwise there are small excitations in the system that grow exponentially in time. In these equations the normal mode frequencies of the two condensate species are coupled by the inter-species interactions and in the absence of the inter-species term, $\chi_{12}=0$, we have two decoupled condensate spectra \eq{physmodes}. The simplest case is obtained when both BECs are in the normal dispersion regime with $p_1,p_2 <\pi/2$ and $\chi_{jk}>0$. In that case the instability condition for the modes $q$ reads \cite{Ruostekoski_Dutton}
\beq
\chi_{12}^2>\chi_{11} \chi_{22}+q^2 L\left(j_2\chi_{11}+j_1\chi_{22}\right)/2+q^4 L^2 j_1 j_2/4 \, .
\label{E19}
\eeq
where $j_k=J_k \cos p_k$. In the normal dispersion regime the instability first sets in for the modes for which $|q|$ is small and therefore $q^2 L\propto 1/L$. In the limit of a large lattice $L\rightarrow \infty$, we then obtain the criterion for instability
\beq
\chi_{12}^2\gtrsim \chi_{11} \chi_{22}\,.\label{phasesep}
\eeq
This stability condition is notably altered if one of the BECs exhibits anomalous dispersion due to superfluid flow \cite{Ruostekoski_Dutton}. If the interaction strengths in \eq{E19} are tuned in such a way that the instability is characterized by a single unstable mode, the two-component system can also be found in a state that is no longer dynamically stable but does not undergo a phase separation \cite{Shrestha}. Instead, the two-species mixture exhibits a periodically appearing and disappearing vector soliton structure.

In this work we only consider initially stationary BECs with the vanishing condensate momenta $p_1=p_2=0$. We also assume that the hopping amplitudes for the two BECs are equal $J_1=J_2=J$ and $\epsilon_{1,q}=\epsilon_{2,q}\equiv \epsilon_{q}=4 J \sin^2 (q/2)$. This simplifies the stability analysis. The eigenvalue system for the linear stability analysis in \eq{E10} can then be expressed as ${\mathfrak M}_q$ given by
\beq
{\mathfrak M}_q=
\left(
\begin{array}{cccc}
 \eta_{1,q} & -{{\Delta}}_{11} & {{\Delta}}_ {12} & -{{\Delta}} _{12} \\
 -{{\Delta}}_{11} & \eta_{1,q} & -{{\Delta}} _{12} & {{\Delta}} _{12} \\
 {{\Delta}} _{12} & -{{\Delta}} _{12} & \eta_{2,q} & -{{\Delta}} _{22} \\
-{ {\Delta}}_{12} & {{\Delta}}_{12} & -{{\Delta}}_{22} & \eta_{2,q}
\end{array}
\right)\label{simpleM}
\eeq
with the definition $\eta_{j,q}= U_{jj} +\epsilon_{q}$. The system exhibits two physical normal mode frequencies
\beq
\Omega^\pm_q \equiv \sqrt{\half(t_2\pm t_3)}\, ,
\label{physmodes}
\eeq
For a positive definite ${\mathfrak M}_q$ these are real indicating dynamical stability of the system. The corresponding dynamically stable eigenvectors $\xi_q$ satisfy the  normalization  condition $\xi_q^\dagger \sigma \xi_q=1$. The eigenvalues $-\Omega^\pm_q$ of \eq{E10} represent unphysical solutions with the corresponding eigenvectors satisfying the negative normalization $\xi_q^\dagger \sigma \xi_q=-1$.

The BEC system becomes dynamically unstable when the normal mode frequencies exhibit nonvanishing imaginary
parts, indicating perturbations that grow exponentially in time. The corresponding eigenvectors satisfy $ \xi_q^\dagger \sigma \xi_q=0$.  The rate at which the instability sets in depends on the magnitude of the
imaginary part of the eigenfrequency.

We can solve the eigenvectors of \eq{E10} analytically. The expressions for the mode functions notably simplify when we consider the case $\chi_{11}=\chi_{22}$. We then obtain (for $\chi_{12}\neq0$)
\begin{eqnarray}
\Omega_q^\pm & = \sqrt{\epsilon_q(\epsilon_q+2\Delta_{11}) \pm 2 \epsilon_q \Delta_{12}}\label{omegasimple}\\
u_{q,\pm}^{(1)} &= { 4 \Delta_{12}\Omega_q^{\pm}+ (\Omega_q^{+})^2- (\Omega_q^{-})^2\over 4 \sqrt{2{\Delta}_{12} \Omega_q^{\pm} \big[ (\Omega_q^{+})^2- (\Omega_q^{-})^2 \big] }}\label{usol}\\
v_{q,\pm}^{(1)} &= {  4 \Delta_{12}\Omega_q^{\pm}+ (\Omega_q^{-})^2- (\Omega_q^{+})^2  \over 4 \sqrt{2\Delta_{12} \Omega_q^{\pm} \big[ (\Omega_q^{+})^2- (\Omega_q^{-})^2 \big] }}\,\label{vsol}
\end{eqnarray}
and $u_{q,\pm}^{(2)}=\pm u_{q,\pm}^{(1)}$, $v_{q,\pm}^{(2)}=\pm v_{q,\pm}^{(1)}$. Here we have assumed, for the notational simplicity, that $\Delta_{12}>0$. In TWA simulation we generate the initial state noise for the configuration that is dynamically stable (we specifically consider thermal equilibrium states). For such states the normal mode frequencies are all real and the eigenmodes satisfy the normalization condition
\beq
\xi_q^\dagger \sigma \xi_q = [u_{q,\pm}^{(1)}]^2 -[v_{q,\pm}^{(1)}]^2 + [u_{q,\pm}^{(2)}]^2 -[v_{q,\pm}^{(2)}]^2=1\,.
\eeq

The two-species BEC normal modes describe the dynamics of mixing between the two components as well as excitations of the total density in the system. In nonlinear regime we have $\chi_{jj}\gg
J$. If we also have $\chi_{12}^2\simeq \chi_{11}\chi_{22}$, one of the frequencies approaches zero corresponding to the phase separation instability \eq{phasesep}. We obtain
in that case $\Omega_{q,+}^2\simeq \nu_{1,q}^2+\nu_{2,q}^2$
and $\Omega_{q,-}^2\ll \nu_{1,q}^2,\nu_{2,q}^2$. The low energy excitations then correspond to the mixing of the two species with only a weak variation in the total density of the two-species condensate.

In both stable and unstable regimes of the two-species mixture we can investigate the degree of overlap between the two species. We define the overlap integral of the wavefunctions for the mixture as
\beq
\kappa(t)=\bigg|\sum_{n}[\psi_n^{(1)}(t)]^*\psi_n^{(2)}(t)\bigg|^2\, .\label{overlap}
\eeq
In the stable regime \eq{overlap} describes the spin excitations of the two-component system.
When the two-species interaction strengths satisfy the condition \eq{E19}, the two-species system is dynamically unstable and undergoes phase separation, resulting in strongly reduced overlap integral values. A measure that can particularly well identify the phase-separation of the densities of the two species may be calculated from the sum
\beq
\tau(t)=\sum_{n}|\psi_n^{(1)}(t)|^2 |\psi_n^{(2)}(t)|^2\,. \label{overlap2}
\eeq

\subsection{Truncated Wigner approximation}

In the TWA simulations we calculate ensemble averages of stochastic trajectories for which the time evolution follows the classical mean-field theory, but in each realization the initial state is stochastically sampled from a Wigner distribution that approximately synthesizes the quantum statistical correlations of the initial state. Approaches introduced in the TWA initial state generation in single-component BECs involve evaluating the initial state correlations within the Bogoliubov approximation \cite{Isella_PRA_2005,Isella_PRA_2006} or by solving the ground state and the excited state populations self-consistently within the Hartree-Fock-Bogoliubov approximation \cite{Gross_PRA_2011}.

In order to implement the TWA phase-space model in a two-component BEC system we similarly assume that the two component stochastic fields $\bar \psi^{(j)}$ obeys the classical field equations similar to the TCDNLSE (Eq.~\ref{E1})
\beq
i{d\over d t}\bar\psi_{n}^{(j)} =-J (\bar\psi_{n+1}^{(j)}+\bar\psi_{n-1}^{(j)})
+\sum_{k=1}^{2}\chi_{jk} |\bar\psi_{n}^{(k)}|^2 \bar\psi_{n}^{(j)}\,.
\label{E30}
\eeq
For the stochastic initial state generation we introduce an approximate model based on the classical Bogoliubov theory, described in the previous section, that provides simple analytically solvable mode functions. We write the both components $j=1,2$ as
\begin{equation}
 \sqrt{N_j} \bar{\psi}^{(j)}_n =  \phi_n^{(j)} \alpha_0^{(j)}+\bar{\delta\psi}_n^{(j)}  \, ,
\label{noise}
\end{equation}
where $\phi_n^{(j)}$ denotes the normalized ground state solution of the BEC component $j$ and the excited-state fluctuations are given by
\beq
\bar{\delta \psi}_n^{(j)} = \frac{1}{\sqrt{L}} \sum_{q\neq0,\eta=\pm} \( u_{q,\eta}^{(j)} {\alpha_{q,\eta}^{(j)}}
e^{i q n }- [v_{q,\eta}^{(j)}]^* [\alpha_{q,\eta}^{(j)}]^*  e^{-i q n} \)
\label{fluctuations1}
\eeq
Here the eigenmodes $(q,\pm)$ correspond to the eigenfrequencies $\Omega_q^\pm$ of \eq{physmodes}.
The mode amplitudes $\alpha_0^{(j)},\alpha_{q,\eta}^{(j)}$ are stochastically sampled from the Wigner distribution of harmonic oscillators as explained below.
We consider a two-species system with equal populations $N_1=N_2\equiv N$, with the interaction strengths satisfying $\chi_{11}=\chi_{22}$. In that case the modes $u_{q,\eta}^{(j)}$ and $v_{q,\eta}^{(j)}$ are
given by Eqs.~(\ref{usol}) and~(\ref{vsol}).

Using the field decomposition \eq{noise} and the mode functions, Eqs.~(\ref{usol}) and~(\ref{vsol}), we then stochastically sample the amplitudes of the mode functions by treating them as ideal harmonic oscillators whose distributions are determined by the corresponding Gaussian Wigner function \cite{Gardiner_Book}
\beq
\label{wigner} W(\alpha_{q,\pm}^{(j)},[\alpha_{q,\pm}^{(j)}]^*)=  \frac{2}{\pi}\tanh \( \xi_q^\pm\) \exp\[ -2|\alpha_{q,\pm}^{(j)}|^2\tanh\( \xi_q^\pm\)\]\,,
\eeq
where $\xi_q^\pm\equiv \Omega_q^\pm /2k_B T$. The stochastic mode function amplitudes $\alpha_{q,\nu}^{(j)}$ produce the ensemble averages
\beq
\langle [\alpha_{q,\pm}^{(j)}]^*\alpha_{q,\pm}^{(j)}\rangle_W =  \bar{n}_{q,\pm}+\frac{1}{2}\, ,
\label{E40}
\eeq
where
\begin{equation}
\bar{n}_{q,\pm}=\frac{1}{\exp(\Omega_q ^\pm/k_B T)-1}\, ,
\label{E41}
\end{equation}
is the usual Bose-Einstein distribution function. The factor $1/2$ in \eq{E40} results from the Wigner distribution that returns symmetrically ordered expectation values, providing the vacuum noise in each mode. For each stochastic realization the total number of excited-state atoms varies according to
\beq
N_{e}^{(j)}=\sum_{q\neq0,\nu=\pm}   \[  \big(|u_{q,\nu}^{(j)}|^2+|v_{q,\nu}^{(j)}|^2\big) \big([\alpha_{q,\nu}^{(j)}]^* \alpha_{q,\nu}^{(j)}-\frac{1}{2}\big) +|v_{q,\nu}^{(j)}|^2 \]
\label{E43}
\eeq
with the average number given by
\begin{equation}
\< N_{e}^{(j)}\> =\sum_{q\neq0,\nu=\pm}\[ (|u_{q,\nu}^{(j)}|^2+|v_{q,\nu}^{(j)}|^2)\bar{n}_{q,\nu}+|v_{q,\nu}^{(j)}|^2 \]
\label{E37}
\end{equation}
The ground state amplitudes $\alpha_0^{(j)}$ fluctuate in each stochastic realization \cite{Martin_PRL_2010,Martin_NJP_2010}. The ground-state atom number is then obtained from the fixed total atom number $N$ in each atomic species, so that in each stochastic realization $N_{c}^{(j)}=N-N_{e}^{(j)}$ and we set $\alpha_{0}^{(j)}=\sqrt{ N_c^{(j)} +1/2}$.

\section{Numerical results}

\subsection{Turning up the optical lattice}

We solve the stochastic dynamics during the turning up of the lattice potential for given initial conditions.
We study the response of the system to the ramping so that both the tunneling coefficients and nonlinearity are time-dependent. We assume that at all times $J_1=J_2=J$ and $\chi_{11}=\chi_{22}$. For simplicity, we consider a situation where the both species have an equal mass $m$, so that the recoil frequencies $\omega_R$ are equal and are given by
\beq
\omega_R=\frac{\hbar\pi^2}{2md^2}\,,
\eeq
where $d$ denotes the lattice spacing. In a deep lattice the hopping amplitude $J$ is then approximately given by \cite{Morsch}
\beq
J=\frac{4 e^{-2 \sqrt{s}} s^{3/4}}{\sqrt{\pi }} \omega_R \, .
\label{E30a}
\eeq
Here $s$ denotes the lattice height in the units of the lattice photon recoil energy. In a tightly-confined elongated 1D trap the atoms are assumed to be confined in the radial vibrational ground state. If the radial confinement is the same for the both species, we obtain
\beq
\chi_{ij} \simeq \sqrt{2 N_i N_j\over\pi} {\Omega_\perp a_{ij}\over l_s}\simeq \sqrt{2\pi N_i N_j}\,{\Omega_\perp a_{ij} s^{1/4}\over d} \,,
\label{interaction}
\eeq
where $\Omega_\perp$ denotes the trapping frequency of the radial confinement and $l_s=(\hbar/m\Omega_s)^{1/2}$, where $\Omega_s\simeq 2 s^{1/2} \omega_R$ is the axial trap frequency at the lattice site minimum.

In our simulations we numerically solve the time-evolution using the split-step method~\cite{JavanainenJPA}.
As the lattice is turned up the hopping amplitude $J$ rapidly decreases according to \eq{E30a} and the interaction strength $\chi_{ij}$ slowly increases according to \eq{interaction}, due to the stronger confinement of atoms in individual sites.
We increase the lattice height linearly at the rate $\delta$, so that the lattice height satisfies $s(t)=s_i+\delta t$, where $s_i$ denotes the initial height. We choose $\delta=2\times 10^{-3}\omega_R$ and $s_i=2$, resulting in the initial value of $J\simeq 0.22 \omega_R$. Unless it is stated explicitly, the length of the lattice is  $L=64$ and the number of atoms in each species in each run is taken to be $N/L=40$. In most cases we choose the initial value for the intra-species interaction strengths $\chi_{11}=\chi_{22} \simeq 0.60 \omega_R$, resulting in $\Omega_\perp a_{jj}/d\simeq 7.9\times 10^{-5}\omega_R$. The initial state fluctuations are evaluated within the Bogoliubov approximation and
the interaction strengths are selected in such a way that the excited state population remains low. For $\chi_{11}=\chi_{22} \simeq 0.60 \omega_R$ and $\chi_{12}/\chi_{11}=0.1$ at $T=0$,
the number of atoms initially depleted from the ground state due to quantum fluctuations $\<N_{e}^{(1)}\>=\<N_{e}^{(2)}\>\simeq 70$ (corresponding to 2.7\% depleted fraction).
All the presented simulation results are for $T=0$, except the finite temperature cases studied in Sec.~\ref{sec:finiteT}.

We study the effect of ramping of the lattice both in the stable $(\chi_{12}\lesssim \sqrt{ \chi_{11}\chi_{22}})$ and unstable $(\chi_{12}\gtrsim \sqrt {\chi_{11}\chi_{22}})$ regime and write the inter-species interaction strength $\chi_{ij}$ (for $i\neq j$) as
\beq
\chi_{ij}=\gamma\sqrt{\chi_{ii}\chi_{jj}}\,.
\label{gamma}
\eeq
The nonlinearity corresponding to the inter-species interactions is tuned by varying the parameter $\gamma$ in \eq{gamma}. Here $\gamma \lesssim 1$ corresponds to the stable regime while $\gamma \gtrsim 1$ implies the dynamical phase-separation instability.
In cases where we study the two-species system in the unstable phase separation regime, we use the initial two-species mixture that is dynamically stable, but change the value of $\gamma$ from the stable to unstable regime immediately after the lattice ramp.

For a fixed nonlinearity quantum fluctuations are enhanced by reducing the atom number (and correspondingly increasing the scattering length). For large atom numbers and in shallow optical lattices quantum effects are weak and the system can be accurately described by classical mean-field theory. As the lattice is turned up, the effective interactions become stronger and quantum fluctuations are enhanced.

The ramping of the lattice is adiabatic if the rate of change of the parameters in the Hamiltonian is slow compared to the lowest collective excitation frequency \cite{JAV99,Isella_PRA_2006}. The fastest varying parameter during the ramping is the hopping amplitude and we require for an adiabatic ramp
\beq
\zeta(t)\equiv \bigg | \frac{1}{J(t)} \frac{\partial J(t)}{\partial t}\bigg |\lesssim {\rm min}[\Omega_q(t)]
\, .
\label{E32}
\eeq
If this condition is not satisfied, the system can be excited from its ground state during the turning up process of the lattice.

\subsubsection{Validity of the parameter regimes}

Our chosen set of parameters, explained in Sec.~\ref{parameters}, captures the essential features of the condensate fragmentation, reduced atom number fluctuations and the domain formation. For the phase separation dynamics, the important condition is that of the dynamical instability $\chi_{12}^2\gtrsim \chi_{11}\chi_{22}$ and the precise ratio $\chi_{11}/\chi_{22}$ is less relevant. Experimentally, the interaction strengths $\chi_{ij}$ can be controlled using two-species Feshbach resonances \cite{Papp} or by introducing a linear electromagnetic coupling between the two internal states \cite{obert11}.
The interaction parameter $\chi_{ij}$ incorporates the atom numbers and the ratio $\chi_{11}/\chi_{22}$ may also be tuned by changing the relative atom population of the two condensate components.
The intra-species interaction strength $\chi_{12}$ can be controlled in a spin-dependent optical lattice \cite{MAN03} by changing the relative lattice positions of the two species and therefore modifying the spatial overlap integral between the lattice site wavefunctions.

The two-species condensate system may be realized by using two different hyperfine levels of the same atom \cite{Hall,Papp} or by trapping two entirely different atoms, e.g., a $^{41}$K--$^{87}$Rb mixture
\cite{CAT07}. In the case of far-detuned optical lattice, the potential experienced by the atoms in two different hyperfine levels of the same atom is typically the same, resulting in the identical values for the hopping amplitudes.

The atom dynamics can be described by a 1D model if the frequency of the radial trapping potential $\Omega_\perp$ is larger than the chemical potential of the atoms $\omega_j$ in Eq.~(\ref{E7}) and the thermal energy $k_B T$. The typical values used in the numerics are $\chi_{11}=\chi_{22}\simeq 0.60\omega_R$ at $s=2$, corresponding to $\chi_{11}\simeq 1.3\omega_R$ at $s=40$ and $\Omega_\perp a_{11}/d\simeq 7.9\times 10^{-5}\omega_R$. We therefore obtain for the requirement of the 1D dynamic description $\Omega_\perp \gtrsim 1.3\omega_R/L$ and $d/a_{11}\gtrsim 250$. The first expression yields in terms of the lattice spacing $d\gtrsim 0.3 l_\perp$ where $l_\perp=\sqrt{\hbar /(m\Omega_\perp)}$ denotes the radial width of the lattice site mode function. For the scattering length $a_{11}\simeq 5$nm we obtain from the second inequality $d\gtrsim 1.3\mu$m. In 1D lattices the lattice spacing can be easily controlled by adjusting the angle between the counter-propagating lasers that form the standing-wave pattern, so that $d=\pi/[k \sin (\theta/2)]$ for the intersection angle $\theta$ and wavenumber $k$. In the simulations we assume 40 atoms per site and we require that the atom density is sufficiently low so that the inelastic atom losses remain weak. The three-body loss rate of the atoms at the site $l$ may be approximated by
\beq
{d n_l\over dt}  = - \Gamma n_l^3, \quad
\Gamma \simeq K_3  \int d^3\vec r\, |\phi_{l}^{(j)}(\vec r)|^6 \simeq  {  K_3 \sqrt{s} m^2 \Omega_\perp^2\over 3\sqrt{3} \pi d^2 \hbar^2}\, ,
\label{3body}
\eeq
where $K_3$ denotes the three-body recombination rate. In order to have a weak three-body loss rate $n_l^3\Gamma /\omega_R \ll 1$, the radial trap frequency should be sufficiently weak. The parameters depend on the particular atom and the hyperfine state. For instance, for $^{87}$Rb in the $|F=1,m_F=-1\rangle$ state we have $K_3\simeq 5.8\times 10^{-30}$ cm$^{6}$/s  \cite{Burt_1997} and we obtain for the condition of the three-body loss rate to be weak $\Omega_\perp \ll 1.2\times 10^5$/s. This condition can be satisfied when the system dynamics is strictly 1D. It should be also noted, however, that even in elongated traps that are not tightly confined, 1D numerical model can provide a good qualitative description of the atom dynamics in the lattice \cite{Gross_PRA_2011}.

In the tight-binding approximation we assume that only the lowest energy band is occupied and one mode function per lattice site is sufficient to represent the dynamics. This approximation becomes better in deeper lattices and provides a reasonable description for $s\gtrsim 2$ \cite{Morsch}. We also require that the nonlinearity is smaller than the energy gap between the lowest two energy bands $\sim \hbar \Omega_s=2\sqrt{s}\, \hbar\omega_R$, so that the higher bands are not occupied. This yields $2\sqrt{s} \omega_R\gtrsim \chi_{ij}/L$ which is well satisfied for all the studied lattice heights.

Experimentally, the atoms are trapped in a combined optical lattice and a harmonic trap. The harmonic potential introduces a non-uniform atom density. This influences the phase separation dynamics; the condensate component with a weaker nonlinearity energetically favours higher density regions close to the centre of the trap \cite{Hall}. The phase coherence and number fluctuations in a harmonic trap depend on the spatial location with quantum and thermal fluctuations stronger close to the edge of the atom cloud \cite{Gross_PRA_2011}. Some other possible effects on the phase coherence are addressed in Sec.~\ref{phasecoherence}. A lattice with a uniform density and periodic boundary conditions may be realized in a toroidal trap with an optical lattice formed by the interference of two counter-rotating Laguerre-Gaussian laser beams \cite{laguerrelattice}.

\subsection{Dynamically stable regime}

We first consider the two-species BEC dynamics in an optical lattice in the dynamically stable regime of the spatially overlapping condensate mixture for $\gamma<1$. This corresponds to the situation where the inter-species interaction is not strong enough to cause the phase separation of the two components and all the normal mode frequencies in \eq{E13} are real.

\subsubsection{Condensate fragmentation and phase coherence}\label{phasecoherence}

The atoms are initially confined in a shallow lattice and we continuously turn up the lattice potential.
The effect of quantum fluctuations on atom dynamics in the lattice can be studied by calculating the phase coherence between the atoms in different lattice sites. When the lattice is turned up the hopping amplitude of atoms between adjacent sites rapidly decreases, resulting in the reduction of kinetic energy of the atoms and hence stronger effective interactions. Quantum fluctuations in the system are enhanced and the phase coherence between the atoms in different sites is destroyed as the condensates undergo fragmentation. We evaluate the loss of phase coherence by calculating the absolute value of the normalized relative intra-species phase coherence between the atoms in different sites,
\beq
C_{k-l}=\frac{\langle{\psi_k^*}^{(j)} \psi_l^{(j)}\rangle}{\sqrt{\langle \psi_k^{(j)}\rangle ~ \langle\psi_l^{(j)}\rangle}},\quad (j=1,2)\, ,\label{intracoherence}
\eeq
between the atoms of the same species $j$ in sites $k$ and $l$. Since we choose $N_1=N_2=N$ and $\chi_{11}=\chi_{22}$, the two species have identical nonlinear properties that are spontaneously broken only due to nonlinear interactions, e.g., in the phase separation. We describe the relative phase coherence between the atoms in the two different species by
\beq
C_{k-l}^{(12)}=\frac{\langle{\psi_k^*}^{(1)} \psi_l^{(2)}\rangle}{\sqrt{\langle\psi_k^{(1)}\rangle ~ \langle\psi_l^{(2)}\rangle}}\, .\label{intercoherence}
\eeq
\begin{figure}
\includegraphics[width=0.9\columnwidth]{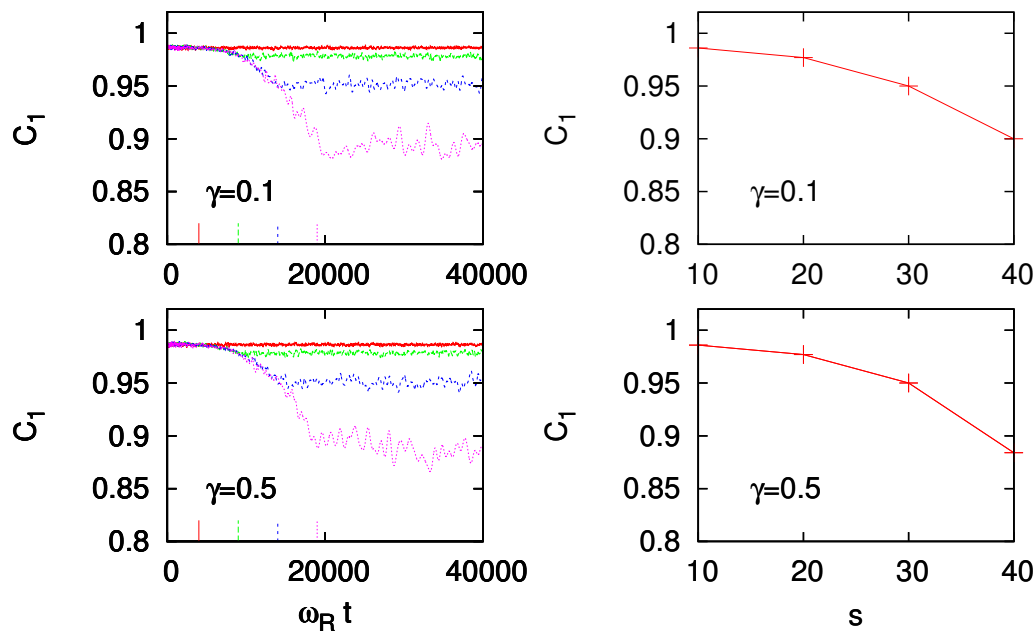}
\caption{Relative phase coherence between the atoms in the same atomic species at two adjacent sites $C_1$ (left) during and after the turning up of the lattice potential for different final lattice heights $s=10,~20,~30,~40$ (curves from top to bottom) with inter-species nonlinear parameter $\gamma=0.1$ (top row) and $\gamma=0.5$ (bottom row). In all the cases the lattice is turned up at the same rate and the end of the ramping times are marked on the $x$ axis using the same colours as in the corresponding curves.
The corresponding stationary (averaged) values of the coherence $C_1$ (right) as a function of lattice height. The coherence $C_1$ monotonically decreases as the lattice is ramped up. The initial lattice height is $s_i=2$ the corresponding nonlinear interaction parameter is $\chi_{11}=0.60\omega_R$. }
\label {f1a}
\end{figure}
\begin{figure}
\includegraphics[width=\columnwidth]{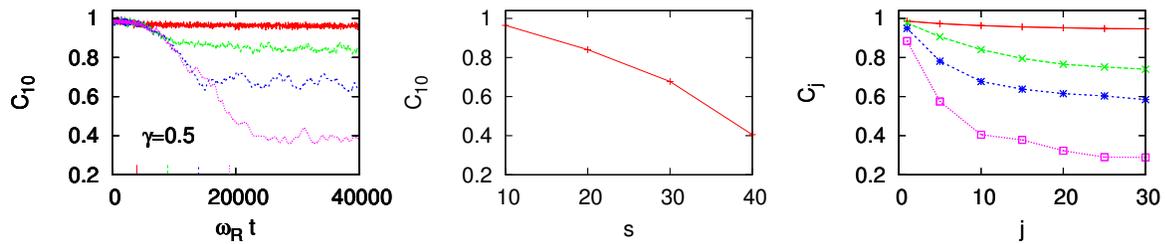}
\caption{Relative phase coherence between the atoms separated by 10 lattice sites and belonging to the same atomic species  $C_{10}$ (left) for different final lattice heights $s=10, ~20,~30,~40$ (curves from top to bottom) with inter-species nonlinear parameter $\gamma=0.5$. The end of the ramping times are marked on the $x$ axis. The corresponding stationary (averaged) values of $C_{10}$ (middle) as a function of lattice height. The spatial dependence of the relative phase coherence along the lattice between the atoms in the same species, displaying the stationary (averaged) values of $C_{j}$ as a function of the relative lattice position $j$ (right).
We specifically show $C_j$ for $j=1,5,10,15,20,25,30$ for different cases of the final lattice height $s=10,20,30,40$ (curves from top to bottom). The coherence $C_{10}$ decreases notably more rapidly than the coherence $C_1$ between the atoms in adjacent sites as a function of the lattice height. As in the case of $C_1$, we find that $C_{10}$ depends only weakly on $\gamma$ in the stable regime for the values we studied from $\gamma=0.1$ to $0.75$.}
\label {f1a2}
\end{figure}

In Fig.~\ref{f1a} we show the relative intra-species phase coherence between the atoms in two adjacent sites $C_1$ during and after the turning up of the lattice potential. The different curves correspond to different values of the final lattice height $s=10,~20,~30$ and $40$. The displayed cases have the inter-species interaction strength $\gamma=0.1$ and $\gamma=0.5$, defined by Eq.~(\ref{gamma}). We also show the corresponding stationary (averaged) values of the coherence $C_1$ that are obtained after the turning up of the lattice potential. These demonstrate how the intra-species coherence rapidly decreases as the lattice becomes deeper, indicating an increasing degree of fragmentation of the BEC as a function of the final lattice height. We only find a very weak dependence of the intra-species relative phase coherence $C_1$ on the inter-species interaction strength $\gamma$ for the values of $\gamma$ in the stable regime we considered in the simulations (from $\gamma=0.1$ to $0.95$). In Ref.~\cite{CAT07} the presence of $^{41}$K was found to lead to lower visibility of the interference fringes of the $^{87}$Rb in the two-species mixture in an optical lattice. In the experiment, however, the interactions increased the atom density of $^{87}$Rb close to the trap centre due to the inhomogeneous trapping potential and drove the system closer to the onset of the Mott insulator transition, as also demonstrated in Ref.~\cite{Hofstetter} where
the adiabaticity and the visibility of interference fringes in loading  of a bosonic mixture to an optical lattice was studied using Gutzwiller mean-field method.

The intra-species long-range spatial coherence is shown in Fig.~\ref{f1a2}. We display the relative phase coherence $C_{10}$ between the atoms in one of the sites and in its 10th nearest neighbour site. This decays notably faster than the coherence $C_1$ between the atoms in the adjacent sites in Fig.~\ref{f1a}. The spatial dependence of the relative phase coherence along the lattice is also shown in Fig.~\ref{f1a2}. We calculated the stationary, averaged values of $C_j$ for different $j$ when the value of the coherence was stabilized after the end of the lattice ramp. The graphs show the decay of the spatial coherence along the lattice. The coherence exhibits a very slow decay for large values of $j$ and remains high for the case of shallow lattices.

The numerically calculated values of the intra-species relative phase coherence may be compared to the analytic estimates obtained for the ground state of the optical lattice system in Appendix. The results for the nearest-neighbour phase coherence for the linearized Bogoliubov theory of the fluctuations in the ground state, displayed in Fig.~\ref{fig:appendix} in Appendix, provide a good agreement with those obtained in the numerical TWA simulations, shown in Fig.~\ref{f1a}. The long-range coherence along the lattice in the TWA numerics, however, decays more rapidly as a function of the spatial separation than in the case of the ground-state calculation.
\begin{figure}
\includegraphics[width=0.9\columnwidth]{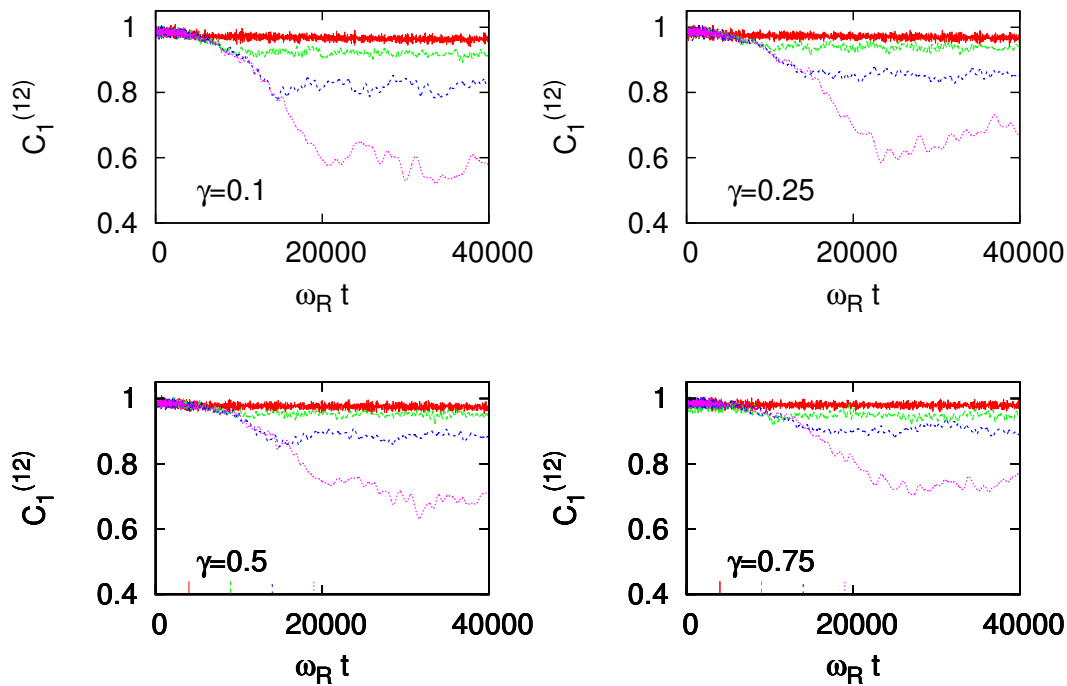}
\caption{Relative phase coherence between the atoms in the different atomic species at two adjacent sites $C^{(12)}_1$ for different final lattice heights $s=10, ~20,~30,~40$ (curves from top to bottom) in the stable regime with inter-species nonlinear parameter $\gamma=0.1,0.25,0.5$ and $0.75$. The other parameters are the same as in Fig.~\ref{f1a}.}
\label {f1b}
\end{figure}

\begin{figure}
\includegraphics[width=\columnwidth]{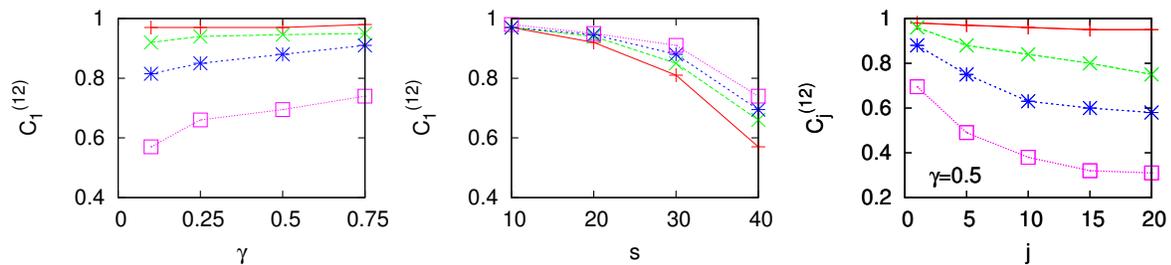}
\caption{The stationary averaged values of the relative inter-species phase coherence $C^{(12)}_1$ as a function of the inter-species interaction strength $\gamma$ (left)  for different values of the final lattice height  $s=10, ~20,~30,~40$ (curves from top to bottom) and as a function of the lattice height (middle) for different values of the inter-species interaction strength $\gamma=0.1,0.25,0.5,0.75$. The stationary values are obtained after the phase coherence is stabilized after the end of the ramping. $C^{(12)}_1$ increases as the inter-species interaction strength increases in the miscible regime. The spatial dependence of the inter-species relative phase coherence along the lattice, displaying the stationary (averaged) values of $C^{(12)}_{j}$ as a function of the relative lattice position $j$ (right).
We show $C^{(12)}_j$ for $j=1,5,10,15,20$ for different cases of the final lattice height $s=10,20,30,40$ (curves from top to bottom).}
\label {f1c}
\end{figure}

The inter-species coherence $C^{(12)}_1$ is shown in Fig.~\ref{f1b} for  different final lattice heights and for different values of inter-species interactions, $\gamma=0.1,~0.25,~0.5$ and $0.75$.
Although the intra-species coherence $C_1$ is not strongly affected by the inter-species interaction strength $\gamma$, the inter-species coherence $C^{(12)}_1$ is very sensitive to $\gamma$ even when the two-species mixture is miscible. In particular, the relative inter-species coherence $C^{(12)}_1$ is enhanced due to the inter-species interactions when $\chi_{12}$ is increased. $C^{(12)}_1$ becomes high as the system approaches to the onset of the phase separation instability, as shown in stationary averaged values of Fig.~\ref{f1c} that are calculated after the turning up of the lattice. This is because the effective interactions in a perfectly overlapping two-species mixture are almost completely canceled out immediately below the onset of the instability for $\chi_{11}\simeq\chi_{22}\simeq\chi_{12}$.
We also show the decay of the spatial coherence along the lattice by displaying $C^{(12)}_j$ for different values of $j$ in Fig.~\ref{f1c} (on right). The stationary, averaged values for the coherence are calculated after the end of the ramp when the coherence has stabilized.

\subsubsection{Atom number fluctuations}

The stochastic simulations in the Wigner representation yield expectation values for the operators that are symmetrically ordered. These can be transformed to normally ordered expectation values of the atoms in each lattice site. For instance, the atom number expectation values and fluctuations are obtained from
\beq
 {\langle n_l^{(j)}\rangle}={\langle [\psi_l^{(j)}]^* \psi_l^{(j)} \rangle -\frac{1}{2} }\, ,
\eeq
\begin{equation}
\Delta n_\idx ^{(j)}=\sqrt{\langle([\psi^{(j)}_\idx]^*\psi^{(j)}_\idx)^2\rangle-\langle[\psi^{(j)}_\idx]^*\psi^{(j)}_\idx\rangle ^2-\frac{1}{4}}\, ,
\end{equation}
for each species-$j$ at the lattice site $l$.

It is useful to scale the atom number fluctuations to those obtained in the Poissonian limit that
correspond to the fluctuations resulting in an instantaneous splitting or in the splitting of a non-interacting gas
\beq
 n_{\rm sqz}^{(j)}=\frac{\Delta n^{(j)}}{\sqrt{\langle n^{(j)} \rangle}}\, ,
\eeq
so that the values $n_{\rm sqz}^{(j)}<1$ indicate reduced on-site atom number fluctuations. The on-site and the relative atom number fluctuations between the atoms in different lattice sites were calculated within TWA for a single-species BEC and compared with experimental observations in Ref.~\cite{Gross_PRA_2011} providing a good qualitative agreement.
\begin{figure}
\includegraphics[width=0.9\columnwidth]{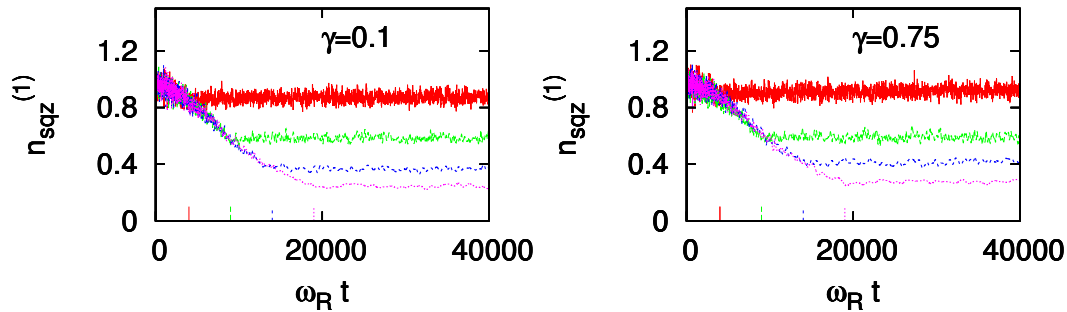}
\caption{Scaled on-site atom number fluctuations $n_{\rm sqz}^{(j)}$ in one of the lattice sites in the stable regime for different final lattice heights $s=10,20,30,40$ during and after the turning up of the lattice for different values of the inter-species nonlinearity $\gamma=0.1$ and $0.75$. The other parameters are same as those in  Fig.~\ref{f1a}. Due to the symmetry $\chi_{11}=\chi_{22}$ the average atom number fluctuations are the same for the both species of the two-component system. }
\label {f2a}
\end{figure}
\begin{figure}
\includegraphics[width=0.9\columnwidth]{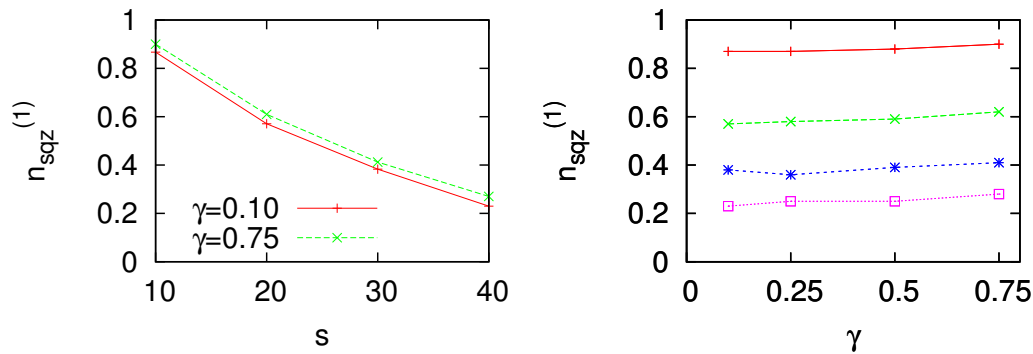}
\caption{The stationary averaged values of the scaled on-site atom number fluctuations $n_{\rm sqz}^{(j)}$ as a function of lattice heights $s$ (left) for different values of inter-species nonlinearity $\gamma=0.1,0.75$ and as a function of the inter-species interaction strength $\gamma$ (right) for different lattice final heights  $s=10, ~20,~30,~40$ (curves from top to bottom). The parameters are the same as in Fig.~\ref{f2a}. The number fluctuations are strongly reduced as a function of the final lattice height, but the dependence of the number fluctuations on the inter-species interactions is less significant.}
\label {f2b}
\end{figure}

In Fig.~\ref{f2a} we show the scaled on-site atom number fluctuations $n_{\rm sqz}$ in one of the sites for the different interaction strengths. The stationary averaged values of atom number fluctuations are shown in Fig.~\ref{f2b}. We show both the dependence of the fluctuations on the lattice height as well as on the inter-species interactions. The atom number fluctuations are strongly reduced as the final lattice height is increased. Strong suppression on number fluctuations for deep lattices is associated with the enhanced phase fluctuations and the condensate fragmentation, and the reduced atom number fluctuations correlate with the previously calculated values of the loss of phase coherence. We also find that the inter-species interactions generally enhance the atom number fluctuations.

The non-equilibrium dynamics of the TWA simulations may again be compared to the analytic estimates obtained for the ground state of the optical lattice system in Appendix. The agreement between the linearized ground-state results of Fig.~\ref{fig:appendix} in Appendix and the TWA results of Figs.~\ref{f2a} and~\ref{f2b} is very good, indicating that the effects of nonadiabaticity in the lattice ramping on the on-site atom number fluctuations are weak.

\subsubsection{Adiabaticity and excitations of relative atom populations}

\begin{figure}
\includegraphics[width=0.9\columnwidth]{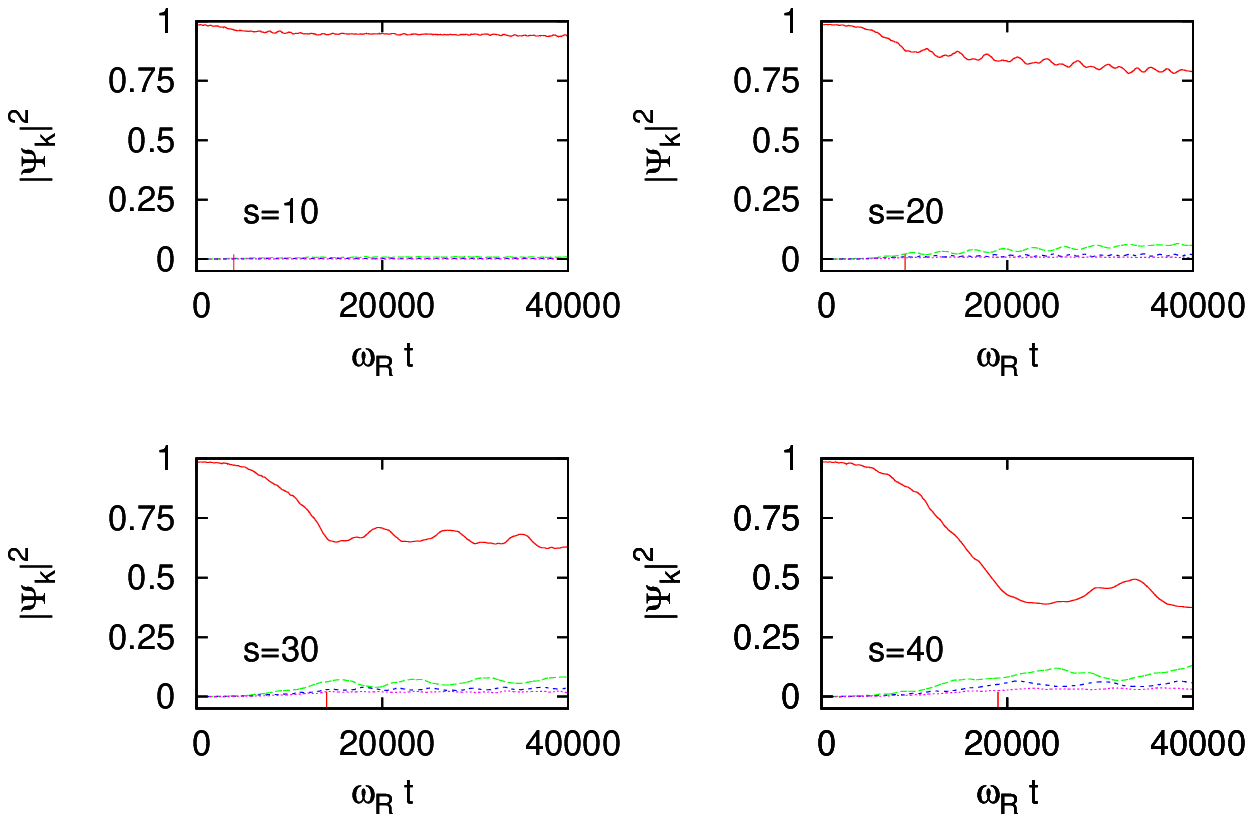}
\includegraphics[width=0.9\columnwidth]{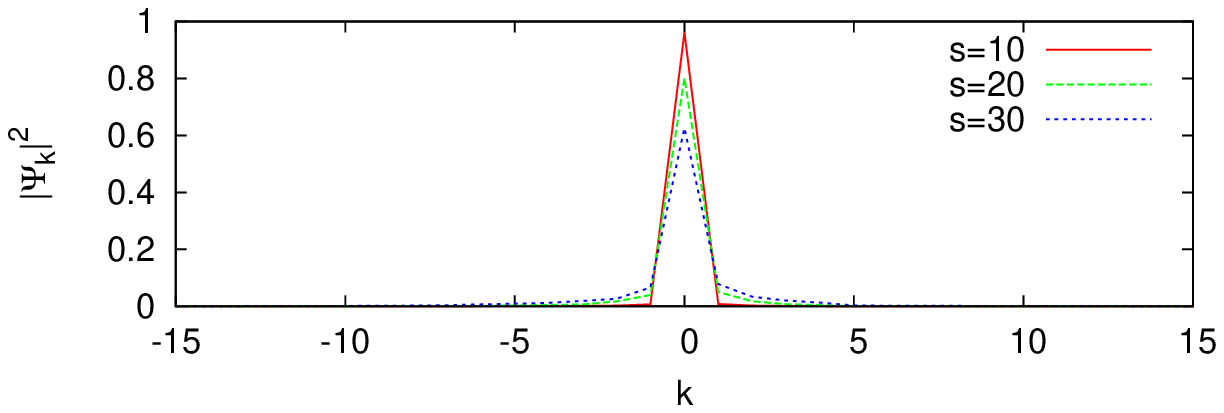}
\caption{The population of first five lowest momentum Fourier modes during the turning up of the lattice potential for different final lattice heights $s=10,20,30,40$ (top two rows). Here $\gamma=0.1$. The lowest mode is initially  occupied while the population of the higher modes is negligible. In the case of a deep lattice the population of excited modes is increased, indicating nonadiabatic turning up of the lattice potential. The ensemble-averaged populations of all the momentum modes at $t=20000/\omega_R$ (bottom row). Only the lowest momentum modes are occupied.}
\label {f3}
\end{figure}

In Fig.~\ref{f3} we show the population of the first five lowest momentum Fourier modes for the inter-species interaction strength $\gamma=0.1$ as a function of time and a snapshot momentum distribution of all the modes. The four figures correspond to the value of lattice height $s=10,~20,~30,~40$. Turning up of the lattice potential results in a decrease in the lowest mode population. The effect is stronger when the lattice becomes deeper, indicating breakdown of the adiabaticity in turning up of the lattice. Consequently, the deeper the lattice, the larger the depletion in the lowest mode. The breakdown of the adiabaticity in deep lattices may be understood from \eq{E32} since the frequency of the lowest phonon mode is reduced as the lattice becomes deeper and it becomes progressively more difficult to turn up the lattice adiabatically.

We may estimate the adiabaticity of the turning up of the lattice potential using the expression \eq{E32}. In the TWA simulations the lattice is turned up at the rate $s(t) = s_i +\delta t$, with $s_i=2$. For $\gamma=0.1$ and $\delta=2.0\times 10^{-3}\omega_R$ we obtain $\Omega_q^{\rm min}(t)/\zeta(t)\simeq 10$ at $s=2$ and 0.7 at the end of the deep lattice ramp $s=40$. The condition  $\Omega_q^{\rm min}(t)/\zeta(t)\simeq 1$ is reached at about $s\simeq 34$. The adiabaticity condition is much easier to violate close to the onset of the phase separation. For $\gamma=0.95$ and $\delta=2.0\times 10^{-3}\omega_R$ we obtain $\Omega_q^{\rm min}(t)/\zeta(t)\simeq 4$ at $s=2$, reducing to 1 at $s\simeq 12$, and to 0.2 at $s=40$.

\begin{figure}
\includegraphics[width=0.9\columnwidth]{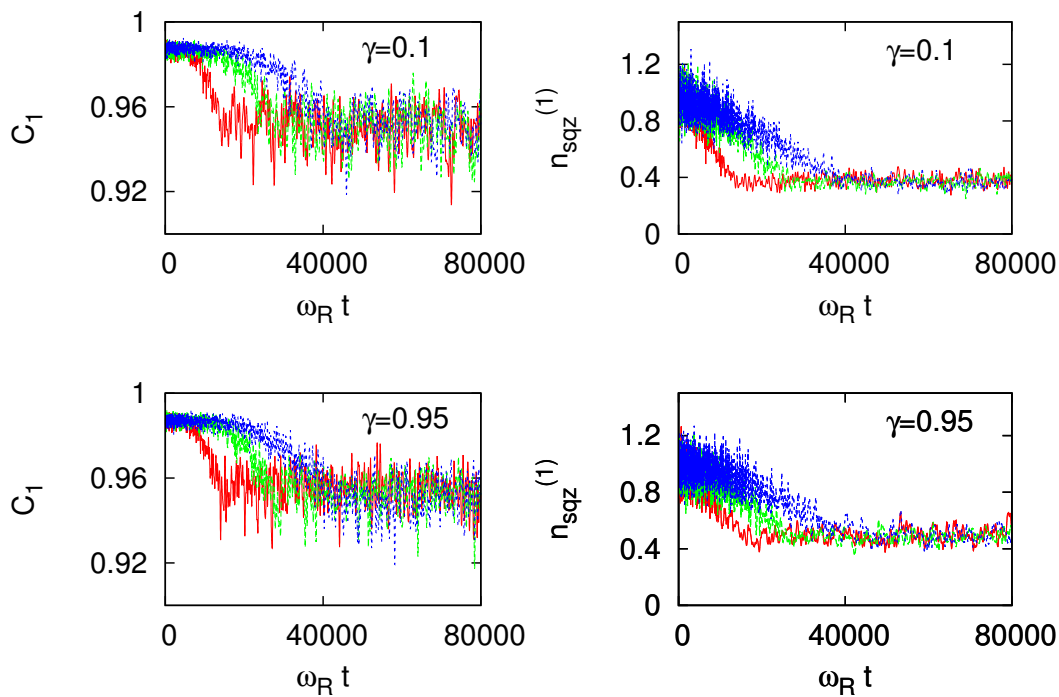}
\caption{The relative phase coherence between the atoms in the adjacent lattice sites $C_1$ and the on-site atom number fluctuations for three different lattice ramp speeds $\delta=2.0\times 10^{-3}\omega_R$, $1.0\times 10^{-3}\omega_R$ and $0.67\times 10^{-3}\omega_R$. The lattice is turned up according to $s(t)=s_i+\delta t$. We show two different cases of $\gamma=0.1$ (top row) and 0.95 (bottom row).}
\label{fig:ramp}
\end{figure}

In Fig.~\ref{fig:ramp} we show the the relative phase coherence between the atoms in the adjacent lattice sites and the on-site atom number fluctuations for three different speeds of the lattice ramp, representing $\delta=2.0\times 10^{-3}\omega_R$, $1.0\times 10^{-3}\omega_R$ and $0.67\times 10^{-3}\omega_R$. For $\gamma=0.1$ and the two slowest ramp cases the condition $\Omega_q^{\rm min}(t)/\zeta(t)\simeq 1$ is never reached during the turning up of the lattice potential. For $0.67\times 10^{-3}\omega_R$ the initial value $\Omega_q^{\rm min}(t)/\zeta(t)\simeq 29$ at $s=2$ is reduced to about 2.1 at $s=40$ for $\gamma=0.1$ and from about 12 at $s=2$ to about 0.5 at $s=40$ for $\gamma=0.95$ ($\Omega_q^{\rm min}(t)/\zeta(t)\simeq 1$ is reached at about $s\simeq 29$).

Despite the improvement in the adiabaticity condition between the three different ramps, there are very little changes in Fig.~\ref{fig:ramp}, especially in atom number fluctuations. Reaching the limit where the ratio $\Omega_q^{\rm min}(t)/\zeta(t)\gg 1$ is both numerically and experimentally demanding in the case of optical lattices with large occupation numbers and a large number of sites. For instance, with the present parameter values, maintaining $\Omega_q^{\rm min}(t)/\zeta(t)\gtrsim 10$ during the entire ramp to $s=40$ for $\gamma=0.95$ would already require a very slow ramp speed of $\delta\simeq 0.2\times 10^{-3}\omega_R$. The difficulty of adiabatically turning up a lattice potential in the experiments can severely limit possibilities to reach the superfluid Mott-insulator transition in the case of large occupation numbers \cite{Isella_PRA_2006}. Achieving a strong reduction in atom number fluctuations in the case of many atoms has been experimentally challenging even in the lattice systems of only a few sites \cite{Esteve_Nature_2008}.

\begin{figure}
\includegraphics[width=0.9\columnwidth]{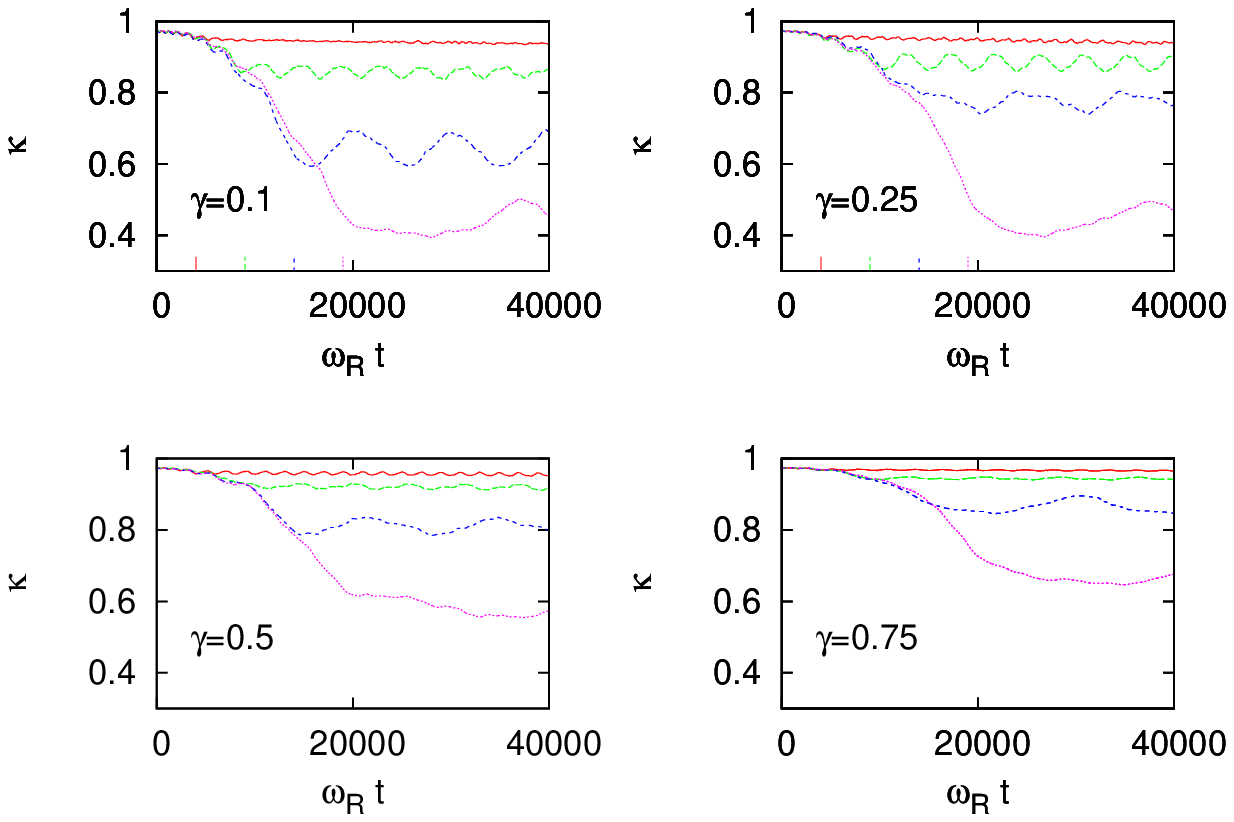}
\caption{Overlap integral between the two condensate species [as defined by \eq{overlap}] in the stable regime for different values of inter-species interactions $\gamma=0.1, 0.25, 0.5, 0.75$ for the final values of the lattice height $s=10,20,30,40$ (curves from top to bottom). }
\label {f4}
\end{figure}

The excitation of higher modes [Fig.~\ref{f3}] induced during the ramping process indicate a nonadiabatic turning up of the lattice. The lattice-induced excitations are also reflected in relative atom populations between the two species (spin-1/2 waves). We show such excitations in the stable regime $\gamma<1$ in Fig.~\ref{f4} by displaying the overlap integral, defined by \eq{overlap}, between the wave functions of two atomic species for various values of inter-species nonlinear parameter $\gamma=0.1,~0.25,~0.5,~0.75$. In deeper lattices the population difference is clearly increased. The effect of interactions, however, is again reduced as $\gamma\rightarrow1$.

\subsubsection{Effects of temperature}\label{sec:finiteT}

\begin{figure}
\includegraphics[width=0.9\columnwidth]{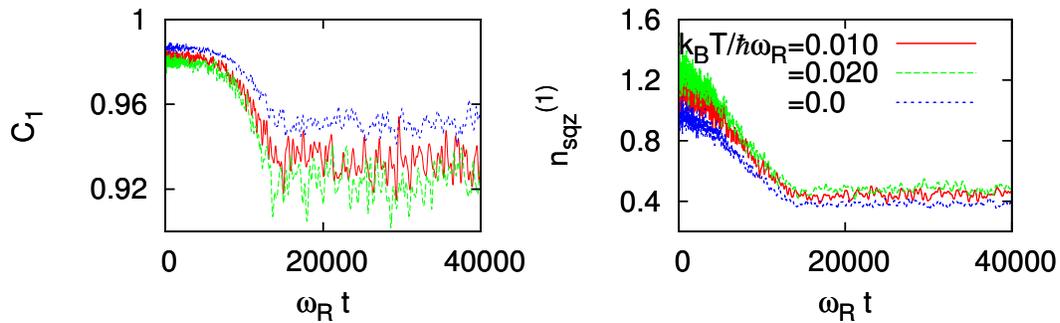}
\caption{Effect of temperature on the on-site atom number fluctuations $n_{\rm sqz}^{(j)}$ and on the relative intra-species phase coherence between the atoms in the adjacent lattice sites $C_1$. Here the nonlinearities $\chi_{11}=\chi_{22}=0.6\omega_R$, the lattice height s=30 and $\gamma=0.50$. The number of excited-state atoms per condensate component for $k_B T/\hbar\omega_R=0.01$ and 0.02 are approximately 370 and 870, respectively. }
\label{f8}
\end{figure}

\begin{figure}
\includegraphics[width=0.9\columnwidth]{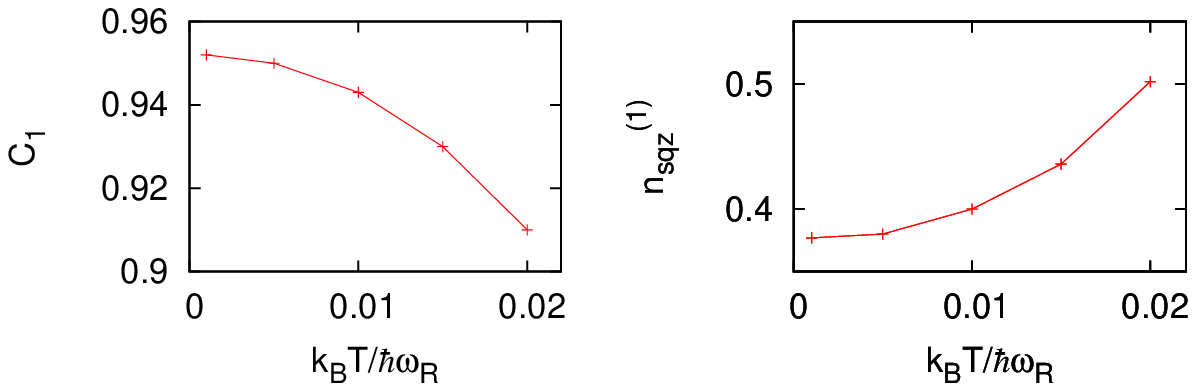}
\caption{The stationary averaged values of the relative intra-species phase coherence between the atoms in the adjacent lattice sites $C_1$ and the on-site atom number fluctuations $n_{\rm sqz}^{(j)}$ as a function of temperature, corresponding to Fig.~\ref{f8}.  }
\label{f8a}
\end{figure}

Finite temperature in non-equilibrium quantum dynamics introduces additional noise in the system, increasing the atom number fluctuations. In experiments on atom number squeezing and reduced on-site atom number fluctuations in optical lattices the finite temperature has been an important factor limiting the achievable spin squeezing and the suppression of the atom number fluctuations \cite{Esteve_Nature_2008,Gross_PRA_2011}.
Here we demonstrate the effects of the initial temperature of the atoms on the coherence properties of the two-species system as the lattice potential is turned up. The temperature can be incorporated in the stochastic sampling of mode populations according to Eq.~(\ref{E41}) in which case the width of the Gaussian stochastic distribution for the sampling of the initial state is increased due to thermal population of each phonon mode. In Fig.~\ref{f8} we show the  effect of variation of temperature on the relative intra-species phase coherence between the atoms in different lattice sites and on the on-site atom number fluctuations. The inter-species interaction parameter $\gamma=0.5$. The corresponding stationary averaged values of atom number fluctuations $n_{\rm sqz}^{(j)}$ and the relative intra-species phase coherence between the atoms in the adjacent lattice sites $C_1$ are shown in Fig.~\ref{f8a}.

\subsection{Unstable regime}

In the previous section we considered a two-species condensate mixture in the regime where the spatial overlap of the two species is dynamically stable, corresponding to the values of the inter-species interaction strength $\gamma\lesssim 1$. When the parameter $\gamma$ is increased the system becomes dynamically unstable and the normal mode frequencies of \eq{E13} exhibit nonvanishing imaginary parts, indicating perturbations that grow exponentially in time. The instability criteria in different regimes for static and moving condensates were analyzed in detail in Refs.~\cite{Ruostekoski_Dutton,Shrestha} and also the effects of matter-wave grating of the other species \cite{Modugno} have been studied. Phase separation is a generic phenomenon that occurs in different forms of matter. The phase separation instability condition for a two-species BEC in a lattice is analogous to the phase separation instability criterion of the two BEC components that occurs in free space when the square of the inter-species interaction coefficient exceeds the product of the intra-species interaction coefficients.

Here we consider the unstable regime of $(\gamma>1)$ by first evaluating the thermal equilibrium state of the atoms in the initial state for some value of $\gamma<1$ corresponding to a stable regime of overlapping two-species mixture. Stable initial ground-state configuration allows us to evaluate the statistical noise for the initial state of the TWA simulations within the Bogoliubov approximation. We then continuously turn up the optical lattice potential as in the dynamically stable case and immediately after the final lattice height is reached we change the inter-species interaction $\gamma$ to the unstable regime. Varying the final lattice height and the atom number provides then information about the dependence of the phase separation dynamics on the lattice parameters and on quantum fluctuations. Experimentally the manipulation of the scattering lengths for BECs in order to drive the system from the stable to unstable regime has been realized for a $^{85}$Rb-$^{87}$Rb condensate mixture using a Feshbach resonance \cite{Papp}. Related experiments on single-component BECs by rapidly changing the scattering length from the stable positive to unstable negative value have generated a condensate collapse and the formation of bright solitons \cite{Khaykovich_Science_2002,Strecker_Nature_2002,Cornish_PRL_2006}. In two-species condensates the effective interaction strengths have been manipulated between stable and unstable regimes using dressed atomic states by electromagnetically-induced Raman transitions between the internal atomic states \cite{obert11}.

\subsubsection{Domain formation}

\begin{figure}
\includegraphics[width=0.9\columnwidth]{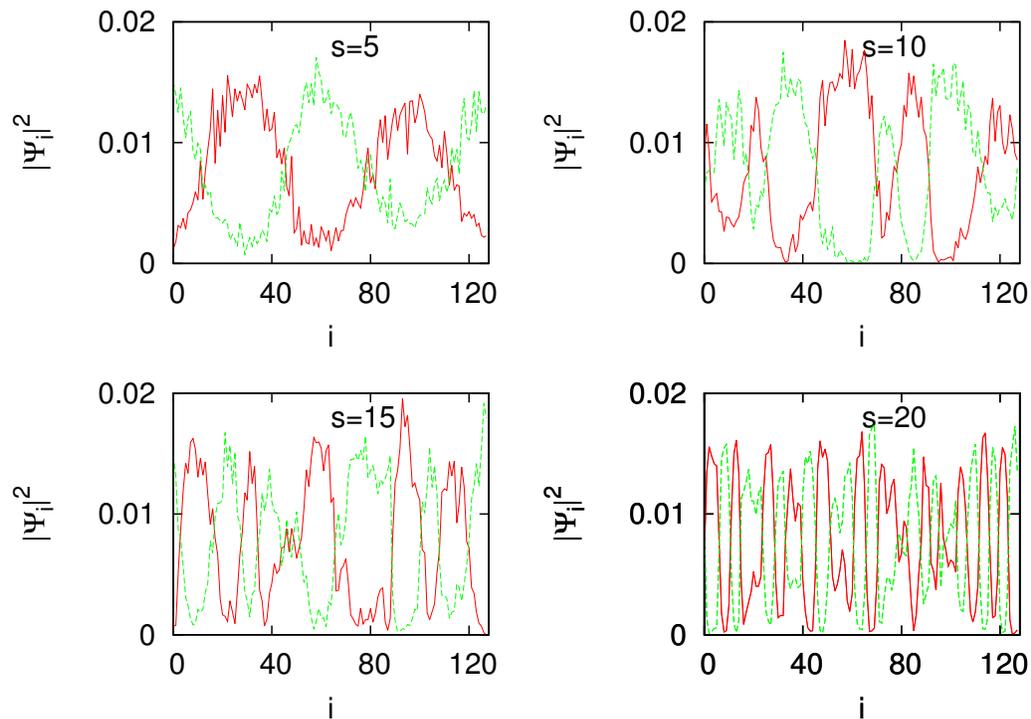}
\caption{Representative individual stochastic realizations of the atom density distributions for the two-species condensates in the dynamically unstable phase separation regime that represent possible outcomes of single experimental runs. The two curves correspond to the densities of the two atomic species. The different figures correspond to different final lattice heights $s=5,~10,~15,~20$ after the system has reached a metastable density configuration. The initial value of the interspecies interaction parameter during the turning up of the lattice potential is in the stable regime with $\gamma=0.95$. After the ramping up of the potential it is changed to the unstable value $\gamma=1.1$. The unstable dynamics results in the spatial phase separation pattern of interleaved density domains. Here the lattice size $L=128$. }
\label{domain_density}
\end{figure}
The system develops instability when the inter-species interaction exceeds a threshold $\gamma \simeq 1$ determined by the intra-species interactions. After the interaction parameters are switched to the unstable regime at the end of the lattice ramping the atom densities of the two BECs show violent phase separation dynamics and individual sites become dominantly occupied by single species alone. In the ground-state configurations of phase-separated systems one species typically forms a shell around the other one, minimizing the surface area between the two components \cite{Ronen}. In an optical lattice system we consider, quantum noise and the nonlinear interactions spontaneously break the symmetry of the uniform spatial configurations of the two species. Individual stochastic realizations of the TWA dynamics that represent possible outcomes of individual experimental runs show density domain formations of the two species. The system settles down to a metastable configuration of several interleaved spatially-separated density domains of the two species where the entire system may consist of multiple domain boundaries along the lattice. Individual sites are typically dominantly occupied by one atomic species alone, except close to the domain boundaries for the case of large domain length. In Fig.~\ref{domain_density} we show typical single realization results for the atom density distribution for lattice heights $s=5,~10,~15,~20$ after the system has phase-separated. The number of domains increases as the lattice height increases.

The observed phenomenon is related to the experiments on a harmonically trapped $^{85}$Rb-$^{87}$Rb condensate mixture in the absence of a lattice \cite{Papp} in which case one of the species was found to split into multiple separated atom cloudlets that appeared as distinct holes in the density distribution of the other species. The condensation experiment was theoretically analyzed in Ref.~\cite{Ronen} and it was demonstrated how the dynamics leads to continuous separation of the two species into smaller and smaller domains.
As shown in Ref.~\cite{Shrestha}, the formation of only a few density spikes and holes in the phase separation dynamics can be identified as a spontaneous generation of bound pairs of dark and bright solitons \cite{Busch_PRL_2001}. The phase separation provides a mechanism for the background of a dark soliton (density hole) in one species to stabilize a bright soliton (density spike) in the other species due to an effective trap that results from the repulsion between the two species \cite{SAV03}. In the present system the stabilization of the small domains is similar to the stabilization in energetically metastable particle-like solitons \cite{SAV03} and results in metastable configurations that are not necessarily energetically close to the ground state. The formation of the metastable states is a non-equilibrium process and the interleaved pattern does not represent a thermal state. One should compare the observed phase-separated state to the ground state that is a maximally phase-separated state and minimizes the surface area between the two components by forcing them to the opposite sides of the trap (one component to the right and the other one to the left).

In the spontaneous pattern formation, due to an instantaneous switch of the interaction strengths, the domain length is expected to be approximately determined by the wavelength of the phonon mode with the largest imaginary part of the eigenfrequency, since this corresponds to the unstable eigenmode that grows most rapidly (provided that the perturbation is strong enough to populate this mode). We can calculate the wavenumber of the mode $q_{\rm max}$ analytically from the expressions of $\Omega_q$.  The value of $q_{\rm max}$ represents the fastest growing mode and gives an order-of-magnitude estimate for the domain length by $l\sim 1/ |q_{\rm max}|$. For $J_1=J_2$ we obtain for the  phase separation  domain length
\begin{eqnarray}
 & |q_{\rm max}|  =|\cos^{-1}\theta |,\quad \theta={\rm max}(-1,\beta)\nonumber\\
 & \beta \equiv \frac{4J + \Delta_{11}+\Delta_{22}-\sqrt{(\Delta_{11}-\Delta_{22})^2 + 4 \Delta_{12}^2}}{4J}=\frac{2J + \Delta_{11}-|\Delta_{12}|}{2J}\, ,
\label{domain}
\end{eqnarray}
where $\Delta_{12}^2 > \Delta_{11} \Delta_{22}$ [where $\Delta_{ij}$ is defined in \eq{E7}] and the latter equality is valid for $\Delta_{11}=\Delta_{22}$.
The domain length from \eq{domain} depends on the lattice height that modifies the dispersion relation. In deep lattices and with stronger interactions the domain length becomes shorter according to \eq{domain}, as also numerically demonstrated in Ref.~\cite{Alon}. Note that reaching the ground state of maximally phase-separated state would require a very slow transition to the unstable regime so that only the lowest energy unstable mode is seeded in the process.
\begin{figure}
\includegraphics[width=0.9\columnwidth]{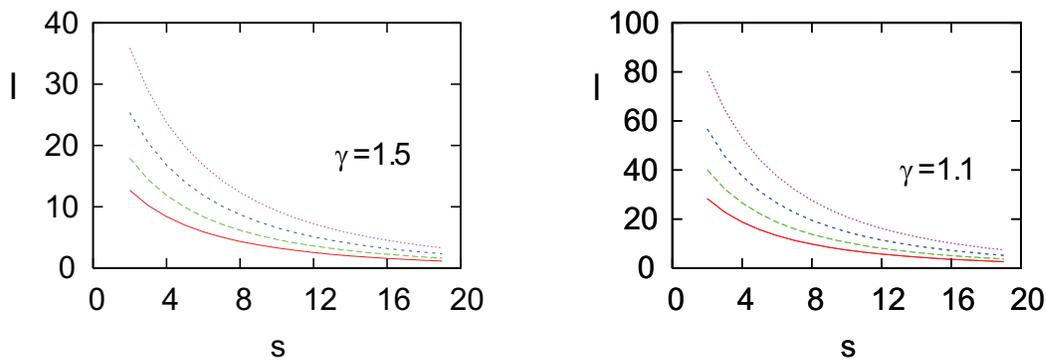}
\caption{Domain length calculated from Eq.~(\ref {domain}) with $l\sim 1/ |q_{\rm max}|$ for different values of the nonlinearity $\Delta_{11}=\Delta_{22}$. The curves from top to bottom represent the values $\Delta_{11}=2.8\times 10^{-3}$, $1.4\times 10^{-3}$, $0.7\times 10^{-3}$, and $3.5\times 10^{-4}\,\omega_R$ at the lattice height $s=2$. The corresponding domain length values for deeper lattices are obtained from the value at $s=2$ by changing $\Delta_{ij}=\chi_{ij}/L$ and $J$ in Eq.~(\ref {domain}) as a function of the lattice height, as in Eqs.~(\ref {E30a}) and~(\ref{interaction}). The different curves can be considered to represent either different atom densities or different interaction strengths. The values of the inter-species interaction parameter $\gamma=1.5$ (left) and 1.1 (right).}
\label {domainlength}
\end{figure}

In Fig.~\ref{domainlength} we show the classical field-theory estimates for the domain lengths obtained from Eq.~(\ref{domain}) with $l\sim 1/ |q_{\rm max}|$ for different values of the nonlinearity as a function of the final lattice depth. The domain length rapidly decreases as the lattice becomes deeper, eventually saturating due to the finite range of available $q$ values in the lowest energy band.

In order to study the effect of quantum fluctuations of the atoms on the domain formation we vary the strength of quantum fluctuations in the simulations. We may continuously interpolate from the regime of strong quantum fluctuations to the classical mean-field limit by keeping the nonlinear interaction strengths $\chi_{ij}$ constant, but varying the atom number \cite{Isella_PRL_2005}. This is done by changing the parameter $\Omega_\perp a_{jj}/( N d)$ for constant $\chi_{ij}$. In the Bogoliubov approximation the nonlinearities $\chi_{ij}$ fix the number of excited-state atoms depleted from the ground state, so varying the total atom number changes the depleted fraction due to quantum fluctuations.
\begin{figure}
\includegraphics[width=0.9\columnwidth]{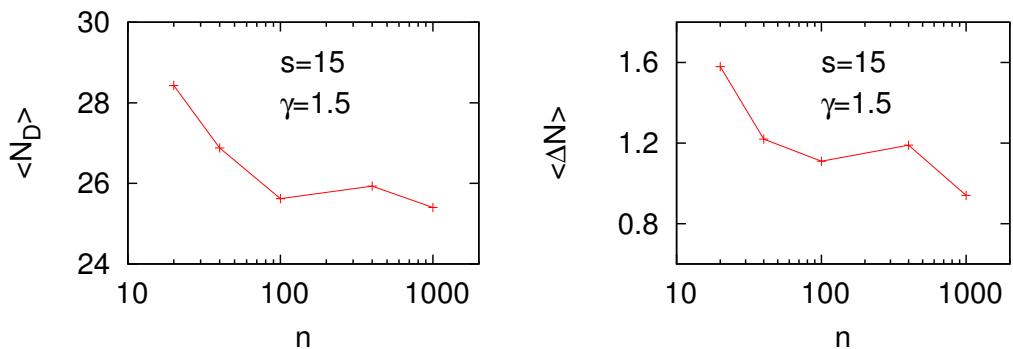}
\caption{Quantum mechanical expectation values and uncertainties for the number of domains in dynamically unstable regime of the two-species condensate in phase separation. We vary the number of atoms per lattice site $n=N/L$ by changing the value of the parameter $\Omega_\perp a_{jj}/( N d)$ while keeping the nonlinear interaction strengths $\chi_{ij}$ constant. The limit $n\rightarrow\infty$ corresponds to the classical mean-field result, while quantum fluctuations become progressively stronger as $n$ is reduced. We find that quantum fluctuations increase the number of domains and the fluctuations in the number of domains. The initial value of the interspecies interaction parameter during the turning up of the lattice potential is in the stable regime with $\gamma=0.95$. After the ramping up of the potential it is changed to the unstable value of $\gamma=1.5$. Here $L=128$ and the final lattice height $s=15$.}
\label {figdomain}
\end{figure}

We show in Fig.~\ref{figdomain} the quantum mechanical expectation value of the number of domains and the corresponding quantum mechanical uncertainty, obtained by the ensemble-averaging stochastic phase-space simulations. The two-species mixture is miscible with $\gamma=0.95$ during the turning up of the lattice potential from $s_i=2$ to $s=15$, after which the system is switched to the unstable regime by changing the parameter value to $\gamma =1.5$. The initial nonlinearity $\chi_{jj}=0.60\omega_R$, but the number of atoms per site $n=N/L$ in one of the species is varied from 20 to 1000, with $L=128$. The limit $n\rightarrow\infty$ corresponds to the classical mean-field limit. We find that both the number of domains and the fluctuations in the number of domains are increased due to quantum fluctuations.

The domain boundaries between the two species may be viewed as defects and by introducing an electromagnetic coupling between the two components that mixes the atom populations provides a phase-separation scheme suitable for testing the Kibble-Zurek mechanism for defect formation in phase transitions \cite{LEE09,Zurek}. Changing the value of the interaction parameter through the phase transition point spontaneously breaks the symmetry of the system in which case the formation of the defects is expected to depend on the timescale of the transition, providing an interesting link to condensed matter physics and cosmology.

\subsubsection{Relative phase coherence and number fluctuations}
	
\begin{figure}
\includegraphics[width=0.9\columnwidth]{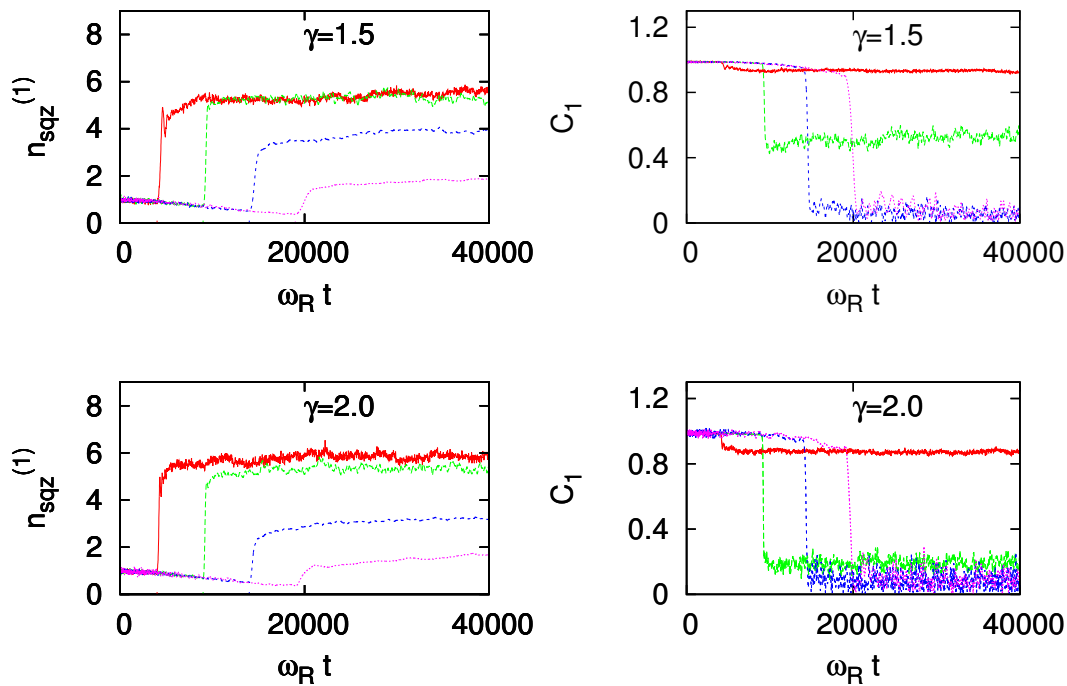}
\caption{On-site atom number fluctuations (left column) and the relative intra-species phase coherence between the atoms in adjacent sites (right column) for one of the atomic species in the unstable regime for different values of the final lattice height $s=10,20,30,40$ during and after the turning up of the lattice.
The initial value of the inter-species interaction parameter during the turning up of the lattice potential is in the stable regime with $\gamma=0.95$. After the ramping up of the potential it is changed to the unstable value of $\gamma=1.5$ (top row) and $\gamma=2.0$ (bottom row). }
\label {f5}
\end{figure}
\begin{figure}
\includegraphics[width=0.98\columnwidth]{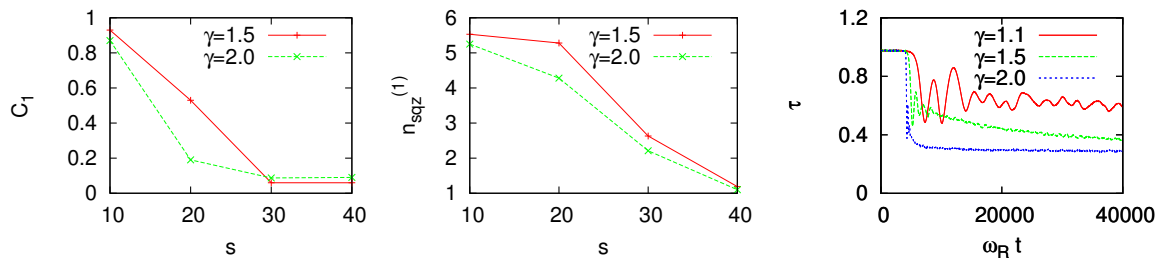}
\caption{Stationary values of fluctuations and the density overlap between the two BEC components. The averaged stationary values of the relative intra-species phase coherence between the atoms in adjacent sites (left) and the on-site atom number fluctuations (middle) in the unstable regime as a function of final lattice height, obtained from the results of Fig.~\ref{f5}. The density overlap integral between the two BEC components $\tau(t)$, defined by \eq{overlap2}, for different values of the inter-species interaction strength $\gamma$. The value depends on the ratio between the interaction strengths $\chi_{11}/\chi_{12}$.}
\label {f5a}
\end{figure}
In the unstable regime there is a dramatic loss in the relative phase coherence between the atoms in different lattice sites and the growth of on-site atom number fluctuations. The enhanced on-site atom number fluctuations may be understood in terms of the domain formation where the spatial location of the domain boundaries fluctuates from one stochastic realization to another.
In Fig.~\ref{f5} we show the atom number fluctuations in one lattice site for one of the atomic species (left column) and the relative intra-species phase coherence between the atoms in adjacent sites (right column) for different lattice heights $s=10,~20,~30,~40$ and different inter-species interactions $\gamma=1.5$ (top row) and $\gamma=2.0$ (bottom row). Even though there is a rapid loss of phase coherence due to the dynamical instability, the nonvanishing value of $C_1$ indicates that even the dynamically unstable system exhibits nonvanishing phase coherence. The loss of phase coherence is much faster due to dynamical instability than due to the ramping of the lattice. The stationary averaged values as a function of the lattice height after the system has reached a metastable configuration are shown in Fig.~\ref{f5a}. We also display in Fig.~\ref{f5a} the density overlap integral between the two BEC components $\tau(t)$ [\eq{overlap2}] for different values of the inter-species interaction strength $\gamma$ that indicates the degree of spatial phase separation between the components.
\begin{figure}
\includegraphics[width=0.9\columnwidth]{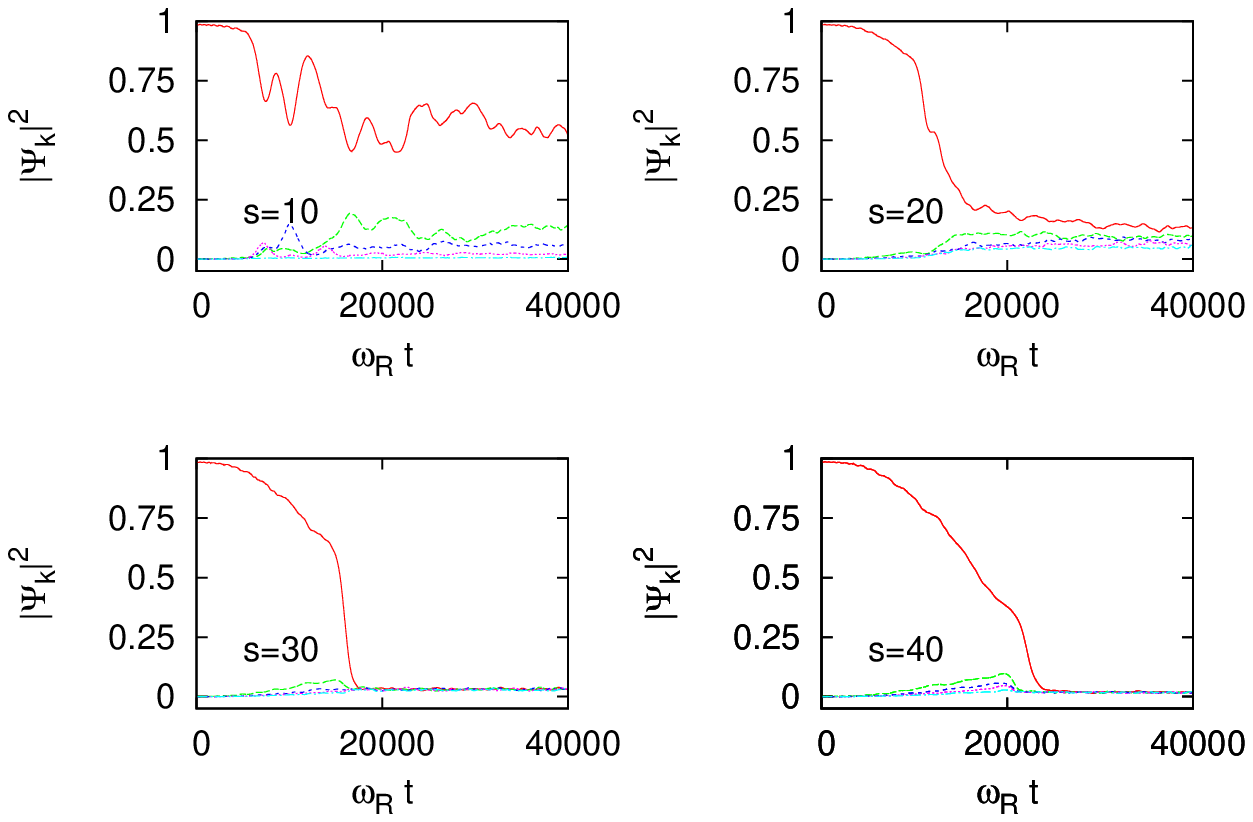}
\includegraphics[width=0.9\columnwidth]{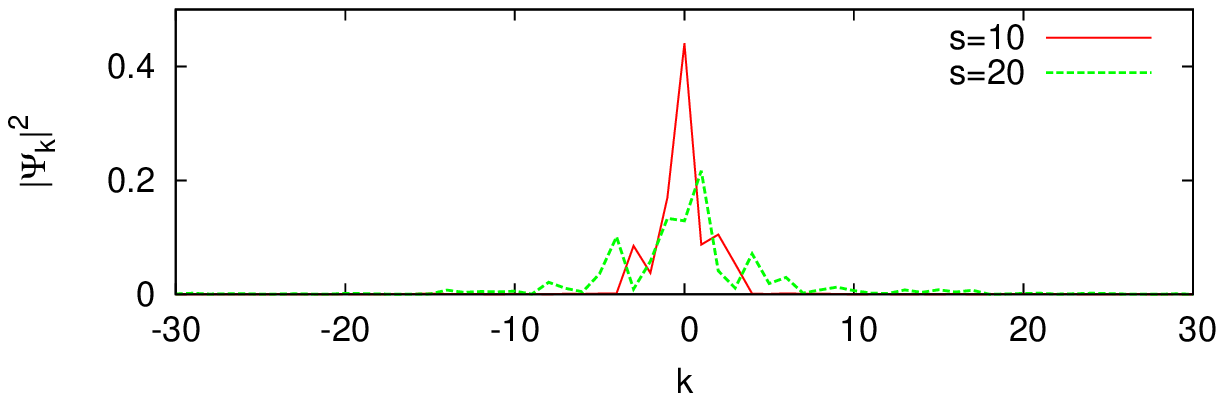}
\includegraphics[width=0.9\columnwidth]{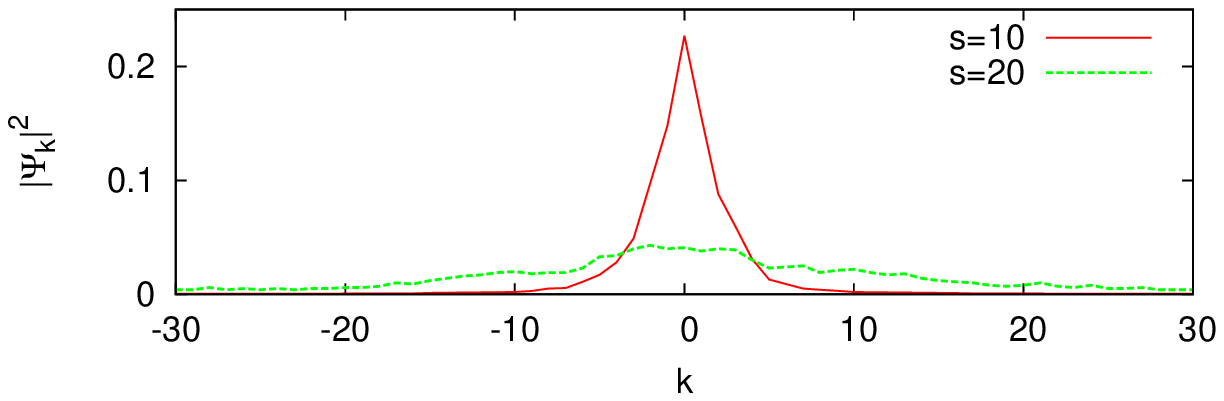}
\caption{Occupations of the lowest five momentum states corresponding to the Fourier modes for different values of the final lattice height $s=10,20,30,40$ in the unstable regime for $\gamma=1.1$, as in Fig.~\ref{f5}. We also show a snapshot image of the occupations of all the momentum components both for a single stochastic realization (middle row) and for the ensemble average over many realizations (bottom row).}
\label {f6}
\end{figure}

In Fig.~\ref{f6} we show the populations in the lowest five momentum modes in the unstable regime. The decay of population in the ground state is fast due to the dynamical instability. The rapid decay of the population is a characteristic feature of the instability. We also show the populations of all the momentum states at a given time ($t=20000/\omega_R$ and $\gamma=1.1$) for a single stochastic realization and for an ensemble average over many realizations.

\section{Concluding remarks}

We demonstrated via numerical simulations how a simple approximate stochastic phase-space method can provide valuable information about a two-species superfluid system in the presence of significant quantum fluctuations. We identified contributions of quantum fluctuations in the pattern formation of the two-species system that results from the modulational instability of dynamically unstable excitations. The parameter space of the two-component condensate system in the lattice is especially large. Although the essential effects of the phase separation dynamics, the loss of the relative phase coherence between the atoms and the reduced atom number fluctuations were captured by the selected parameter regimes, the parameter space could be explored in more detail. In particular, novel phenomena would be observed in the case of moving condensates. The centre-of-mass motion of the atoms results in dissipative transport properties \cite{Fertig_PRL_2005,Isella_PRL_2005}. The stability criteria are also changed when the velocities of the two BECs are different and one of the BECs is not in the normal dispersion regime \cite{Ruostekoski_Dutton}.

Moving condensates may be experimentally studied in a combined optical lattice and a harmonic trap by suddenly displacing the harmonic trap, e.g., by using a magnetic field gradient, in which case the displacement excites dipolar oscillations of atoms along the lattice direction \cite{Fertig_PRL_2005}. The other alternatives to create an analogous effect, e.g., are to use a moving-standing wave, so that the atoms experience a moving optical lattice potential \cite{SAR05} introduce a phase shift for the hopping amplitudes of the atoms between adjacent sites, or to make the hopping amplitudes time-dependent by a periodically pulsating lattice.

It would be particularly intriguing to investigate soliton structures in a two-species BEC. Modifying the ratio $\chi_{22}/\chi_{11}$ \cite{Shrestha}, e.g., by unbalanced relative populations, or inducing flow instabilities \cite{Hamner,Hoefer} would lead to the emergence of vector solitons. The solitons can be persistent (metastable) or pulsating. Another potential application follows from the observation that
repulsive interactions and quantum fluctuations lead to suppressed atom number fluctuations in the lattice. Two-species systems have already been used in experimental realizations of spin-squeezed atom interferometry \cite{Gross_Nature_2010} and squeezing can be employed in interferometers to reach sub-shot-noise accuracies that are not achievable using classical interferometers limited by the standard quantum limit \cite{Holland,Wineland,Bouyer_PRA_1997,Giovannetti_Science_2004}.

\section*{Acknowledgements}

This work formed a part of the project ``Ultracold atomic gases in optical lattices" (10/2010-09/2011) funded by the Leverhulme Trust. We also acknowledge financial support from the EPSRC and the EU STREP NAMEQUAM project.

\appendix

\section*{Appendix: Analytic estimates of atom number and relative phase fluctuations}

\setcounter{section}{1}

Here we provide analytic estimates for the fluctuations of the intra-species on-site atom number and the relative phase between the atoms in different lattice sites in the ground-state of an optical lattice. We introduce the atom number and phase operators for each site as in Ref.~\cite{Isella_PRA_2006}. In the calculations we use the classical Bogoliubov mode functions for the two-species BEC that are employed in the initial state decomposition of the stochastic field in the TWA simulations. By means of replacing the stochastic mode amplitudes $\alpha_{q,\pm}^{(j)}$ in Eq.~(\ref{fluctuations1}) by the annihilation and creation operators $(\hat\alpha_{q,\pm}^{(j)}, [\hat\alpha_{q,\pm}^{(j)}]^\dagger)$, so that
$\alpha_{q,\pm}^{(j)}\rightarrow \hat\alpha_{q,\pm}^{(j)}$ and $[\alpha_{q,\pm}^{(j)}]^*\rightarrow [\hat\alpha_{q,\pm}^{(j)}]^\dagger$, we may obtain analytic estimates for the ground-state properties, provided that the inter-species correlations can be ignored.

We may derive the atom number operator $\hat n_l^{(j)}$ at the site $l$ of the species $j$ by expanding the atom population $n_l^{(j)}$ of the species $j$ in the site $l$ to first order in fluctuation terms. We obtain
\beq
n_l^{(j)} = \( \sqrt{n^{(j)}_{\rm gr}} + \hat{\delta\psi}_l^{(j)} \) \( \sqrt{n^{(j)}_{\rm gr}} + [\hat{\delta\psi}_l^{(j)}]^\dagger \)\simeq {n^{(j)}_{\rm gr}} + \hat n_l^{(j)}\,,
\eeq
where
\beq
\hat n_l^{(j)} = \sqrt{n^{(j)}_{\rm gr}}  \( \hat{\delta\psi}_l^{(j)} + [\hat{\delta\psi}_l^{(j)}]^\dagger \)
=\sqrt{n^{(j)}_{\rm gr}} \sum_{q,\eta=\pm} \(w_{q,\eta}^{(j)} \hat\alpha_{q,\eta}^{(j)} e^{iql} + {\rm H.c.} \)\,,
\eeq
$n^{(j)}_{\rm gr}\simeq N_j/L$ denotes the ground state atom number per site of the species $j$ and $w_{q,\eta}^{(j)} \equiv u_{q,\eta}^{(j)}-v_{q,\eta}^{(j)}$.

We may introduce the corresponding phase operator
at the site $l$ as
\beq
\hat \varphi_l^{(j)}= -\frac{i}{2 \sqrt{N_j}} \sum_{q,\eta=\pm} \(r_{q,\eta}^{(j)} \hat\alpha_{q,\eta}^{(j)} e^{iql} - {\rm H.c.} \)\,,
\eeq
for which the commutator $[\hat n_l^{(j)} , \hat \varphi_l^{(j)} ]=i$ and we have
defined  $r_{q,\eta}^{(j)} \equiv u_{q,\eta}^{(j)}+v_{q,\eta}^{(j)}$.
Then the on-site atom number fluctuations in the $l$th site $(\Delta n_l^{(j)})^2$ and the relative phase fluctuations between the atoms in the $k$th and $l$th site $(\Delta\varphi_{kl}^{(j)})^2 $, respectively, read
\begin{eqnarray}
\label{deltanbogo} (\Delta n_l^{(j)})^2  & = \< [\hat n_l^{(j)}]^2\> - \<\hat n_l^{(j)}\>^2= {n^{(j)}_{\rm gr}\over L} \sum_{q,\eta=\pm} |w_{q,\eta}^{(j)}|^2
(2\bar{n}_{q,\eta}+1)\nonumber\\ & = {n^{(j)}_{\rm gr}\over L} \sum_{q} {\epsilon_q\over 2} \( {2\bar{n}_{q,-}+1\over\Omega_q^-} +{2\bar{n}_{q,+}+1\over\Omega_q^+} \)\,,\\ \label{deltaphibogo}
(\Delta\varphi_{kl}^{(j)})^2  & \equiv
\<(\hat\varphi_k^{(j)}-\hat\varphi_l^{(j)})^2\>= \frac{1}{N_j}\sum_{q,\eta=\pm} |r_{q,\eta}^{(j)}|^2 \sin^2 \[ {q(k-l)\over 2}\] (2\bar{n}_{q,\eta}+1)\nonumber\\ & = {1\over N_j} \sum_{q} {1\over 2\epsilon_q} \[ (2\bar{n}_{q,-}+1)\Omega_q^- +(2\bar{n}_{q,+}+1)\Omega_q^+ \] \sin^2 \[ {q(k-l)\over 2}\]\,,
\end{eqnarray}
where $\bar{n}_{q,\pm}$ is the thermal population of the phonon mode $(q,\pm)$ in the lattice given in Eq.~(\ref{E41}). In the second line of Eqs.~(\ref{deltanbogo}) and~(\ref{deltaphibogo}) we have used the specific relations for $u_{q,\eta}^{(j)}$, $v_{q,\eta}^{(j)}$ and $\Omega_q^{\pm}$ given by Eqs.~(\ref{usol}), (\ref{vsol}) and~(\ref{omegasimple}), respectively. These were obtained by choosing a particular set of parameter values ($\chi_{11}=\chi_{22}$, $J_1=J_2=J$, $N_1=N_2$, etc.). The single-particle energy $\epsilon_q=4 J \sin^2 (q/2)$.

At $T=0$ we have $\bar{n}_{q,\eta}=0$ and we can evaluate the expressions analytically by replacing in the continuum limit the momentum sums by integrals
$${1\over L} \sum_q \rightarrow {1\over 2\pi} \int_{-\pi}^{\pi} dq.$$
For the atom number fluctuations we obtain
\beq
(\Delta n_l^{(j)})^2={n^{(j)}_{\rm gr}\over \pi} \[ \arctan \(\lambda_+\) + \arctan \(\ \lambda_-\) \]\,,
\label{analytic1}
\eeq
where ($\Delta_{11}=\Delta_{22}>\Delta_{12}>0$; $\Delta_{ij}$ is defined in \eq{E7})
\beq
\lambda_\pm=\sqrt{2J\over \Delta_{11}\pm \Delta_{12}}\,.
\eeq
In the nonlinear limit of $(\Delta_{11}\pm \Delta_{12})\gg J$ this simplifies to
\beq
(\Delta n_l^{(j)})^2 \simeq {n^{(j)}_{\rm gr}\sqrt{2J}\over \pi} \( {1\over \sqrt{\Delta_{11}+ \Delta_{12}}} + {1\over \sqrt{\Delta_{11}- \Delta_{12}}} \)\,.
\eeq
Similarly, for the relative phase fluctuations between the atoms in the adjacent sites $(k-l=1)$ we obtain
\beq
(\Delta\varphi_{l, l+1}^{(j)})^2  = {1\over 2 n^{(j)}_{\rm gr} \pi}\[{1\over\lambda_+} +{1\over\lambda_-} + \({1\over\lambda_+^2}+1\) \arctan(\lambda_+) +\({1\over\lambda_-^2}+1\) \arctan(\lambda_-)  \]\,.
\label{analytic2}
\eeq
In the nonlinear case $(\Delta_{11}\pm \Delta_{12})\gg J$ this reads
\beq
(\Delta\varphi_{l, l+1}^{(j)})^2  = {1\over n^{(j)}_{\rm gr} \pi \sqrt{2J}} \( \sqrt{\Delta_{11}+ \Delta_{12}}+ \sqrt{\Delta_{11}- \Delta_{12}} \)\,.
\eeq
Similar relationships as Eq.~(\ref{analytic2}) may also be calculated for different values of $k$ in $(\Delta\varphi_{l, k}^{(j)})^2$ in the continuum limit.
\begin{figure}
\includegraphics[width=0.32\columnwidth]{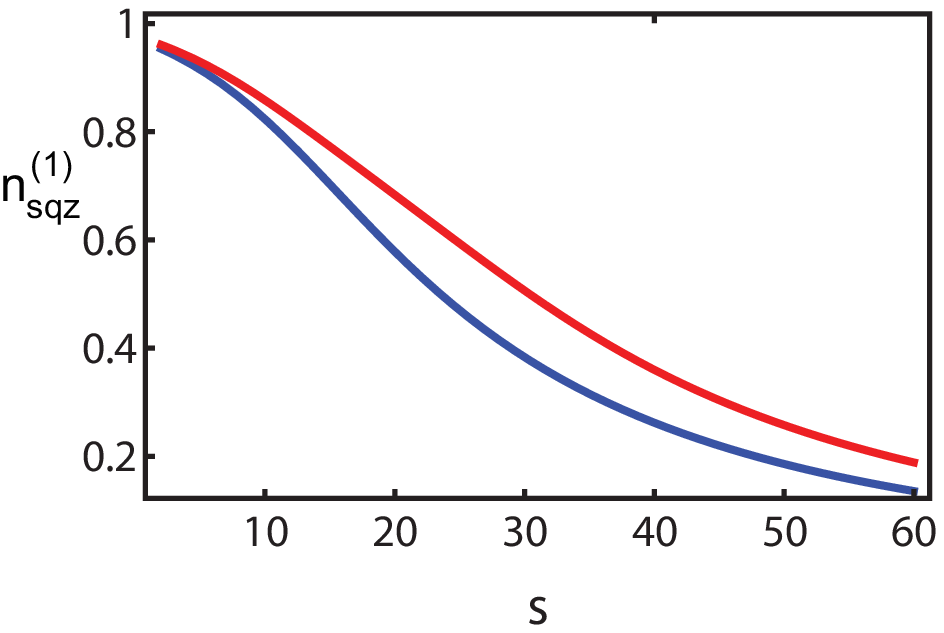}
\includegraphics[width=0.32\columnwidth]{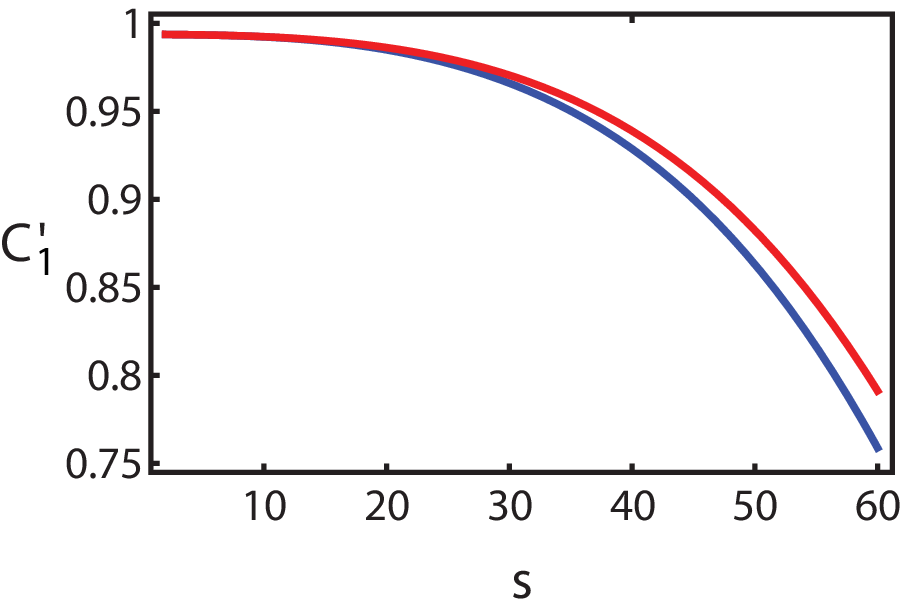}
\includegraphics[width=0.32\columnwidth]{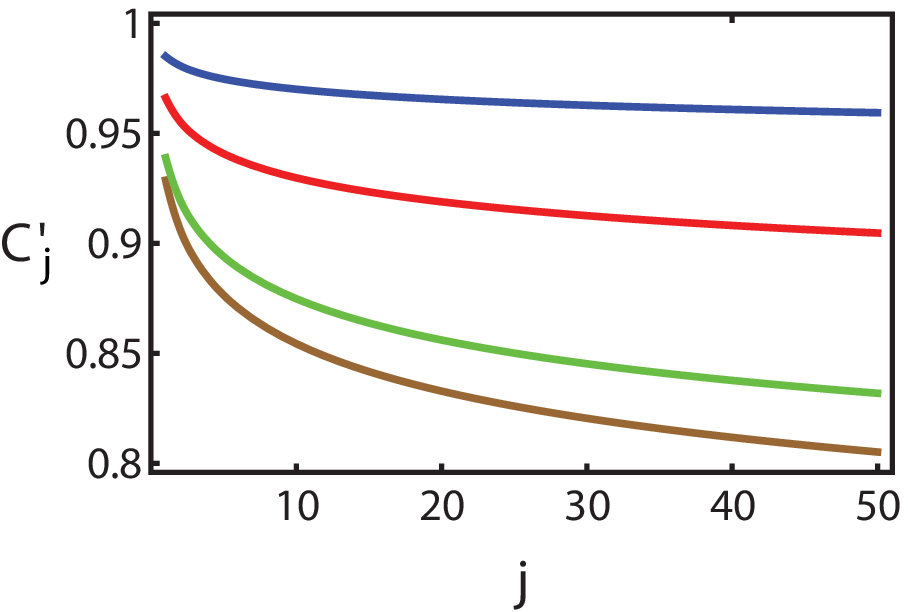}
\caption{The analytic estimates of the ground-state fluctuations. On left, the scaled on-site atom number fluctuations $n_{\rm sqz}^{(1)}=(\Delta n_l^{(j)})/\sqrt{n_l}$ in one of the lattice sites, obtained from Eq.~(\ref{analytic1}). In the middle, the relative phase coherence between the atoms in the adjacent sites $C'_1$, calculated from Eq.~(\ref{phaseanother}). In the both cases the upper curve corresponds to $\gamma=0.9$ and the lower curve to $\gamma=0.1$. On right, we show the phase coherence $C'_j$ along the lattice as a function of the lattice site separation $j$, calculated from Eq.~(\ref{phaseanother}). The curves from the top represent $(s=20,\gamma=0.1)$, $(s=30,\gamma=0.1)$, $(s=40,\gamma=0.9)$ and $(s=40,\gamma=0.1)$.  }
\label {fig:appendix}
\end{figure}

The analytic expression (\ref{analytic2}) for the relative phase fluctuations can be used to evaluate
\beq
C'_l = \left\< \exp \[i\(\hat\varphi_{k+l}-\hat\varphi_{k}\)\] \right\> \simeq \exp \[- \left\< (\hat\varphi_{k+l}-\hat\varphi_{k})^2 \right\>/2 \] \,, \label{phaseanother}
\eeq
as displayed in Fig.~\ref{fig:appendix} in which case we show the relative phase coherence between the atoms in the adjacent sites $C'_1$ together with the on-site atom number fluctuations. We also use the continuum limit approximation to calculate $\left\< (\hat\varphi_{k+j}-\hat\varphi_{k})^2 \right\>$ for different values of $j$ in order to obtain the coherence along the lattice $C'_j$ as a function of the site separation $j$. The parameters of Fig.~\ref{fig:appendix} are the same as those used in in the TWA simulations for the dynamically stable $T=0$ cases, with the dependence of the hopping amplitude $J$ and the nonlinearity $\chi_{11}=\chi_{22}$ on the lattice height determined by Eqs.~(\ref{E30a}) and~(\ref{interaction}), respectively, where we set the value $\chi_{11}=0.6\omega_R$ at $s=2$. The atom number $N=2560$ and the number of sites $L=64$.

In Fig.~\ref{fig:appendix} the relative phase coherence between the atoms in the adjacent sites decreases rapidly as a function of the lattice depth and increases as the inter-species interaction strength $\gamma$ is increased closer to the onset of the phase-separation instability. The numerical values of the nearest-neighbour coherence are very close to those of the TWA simulations in Fig.~\ref{f1a}, but the long-range coherence values are higher than in the TWA case [Fig.~\ref{f1a2}]. One should note, however, that $C_1'$ does not include the atom number contributions incorporated in the definition of $C_1$ [Eq.~(\ref{intracoherence})] which is used in analyzing the relative phase coherence in the TWA numerics.

We can implement a nonlinear least square fit for the coherence along the lattice $C'_j$ using a trial function
\beq
C'_x= a_1 \exp (-a_2 x^{a_3}) +a_4\,,
\eeq
and determine the coefficients $a_i$. An accurate fit for the calculated values in Fig.~\ref{fig:appendix} with $\gamma=0.1$ is obtained for $(a_1,a_2,a_3,a_4)=(0.11,0.55,0.16,0.92)$ for $s=20$ and $(0.24,0.44,0.38,0.77)$ for $s=40$. These indicate asymptotic values for the coherence for large spatial separations 0.92 and 0.77, respectively.

The on-site atom number fluctuations from the analytic estimates in Fig.~\ref{fig:appendix} are also very close to the numerical TWA simulation results of Figs.~\ref{f2a} and~\ref{f2b}. The on-site atom number fluctuations are enhanced as the inter-species interaction strength increases.

\end{document}